# Lost or found? Discovering data needed for research

*Kathleen Gregory*
Data Archiving and Networked Services, Royal Netherlands Academy of Arts and Sciences, Anna van Saksenlaan 51, 2593 HW, The Hague, The Netherlands. E-mail: kathleen.gregory@dans.knaw.nl

*Paul Groth*
Informatics Institute, University of Amsterdam, Science Park 904, 1098 XH Amsterdam, The Netherlands. E-mail: p.groth@uva.nl

*Andrea Scharnhorst*
Data Archiving and Networked Services, Royal Netherlands Academy of Arts and Sciences, Anna van Saksenlaan 51, 2593 HW, The Hague, The Netherlands. E-mail: andrea.scharnhorst@dans.knaw.nl

*Sally Wyatt*
Faculty of Arts and Social Sciences, Maastricht University, Grote Gracht 82, SZ Maastricht, 6211, The Netherlands. E-mail: sally.wyatt@maastrichtuniversity.nl

30 March 2020

## Abstract

Finding data is a necessary precursor to being able to reuse data, although relatively little large-scale empirical evidence exists about how researchers discover, make sense of and (re)use data for research. This study presents evidence from the largest known survey investigating how researchers discover and use data that they do not create themselves. We examine the data needs and discovery strategies of respondents, propose a typology for data reuse and probe the role of social interactions and literature search in data discovery. We consider how data communities can be conceptualized according to data uses and propose practical applications of our findings for designers of data discovery systems and repositories. Specifically, we consider how to design for a diversity of practices, how *communities of use* can serve as an entry point for design and the role of metadata in supporting both sensemaking and social interactions.

## 1. Introduction

Stakeholders from funders to researchers are increasingly concerned with the sharing and reuse of research data **(**e.g. Digital Curation Centre, n.d.; Tenopir et al., 2015**).** Policy makers draft guidelines, systems designers create repositories and tools, and librarians develop training materials to encourage opening and sharing data, often without empirical evidence about community-specific practices (Noorman, Wessel, Sveinsdottir, & Wyatt, 2018). It is assumed that data can and will be reused if they are shared (Borgman, 2015a). Another assumption predicating reuse is that data will actually be discovered by researchers, although relatively little empirical work exists to support this assumption.

In this article, we present the results of the largest known survey examining how researchers discover and (re)use research data that they do not create themselves, so-called *secondary data* (Allen, 2017). We consider commonalities in practices but also examine differences, looking at how data needs and search practices vary not only by disciplinary domain but also by types of data uses. Past work has explored data search practices via in-depth interviews (Koesten, Kacprzak, Tennison, & Simperl, 2017; Borgman, Scharnhorst, & Golshan, 2019; Gregory, et al, 2019a). This study employs a broader approach, using a globally distributed multidisciplinary survey, with nearly 1700 respondents, to explore these practices at a larger scale.

Our study produced a rich dataset including both qualitative and quantitative data. Here, we present the discovery phase of exploring the quantitative data, relying on descriptive statistics and tests for pairwise correlations, and we draw deeply on the qualitative responses. We use these analyses to paint a detailed picture of data discovery and propose a typology for data (re)use which we use to explore the data needs and practices of participants. We also probe the role of social interactions in searching for data and explore how data search is related to other practices, such as searching for academic literature. We consider how communities of data seekers can be conceptualized and discuss our findings in light of recent efforts to increase the discoverability of research data in a theoretical discussion. We conclude with practical considerations for the application of our findings by data discovery systems designers and repository managers and suggest further directions for research.

## 2. Background

Although information seeking is an extensive research field, work investigating *data-seeking practices* is nascent. Practices of data seeking, which we refer to here also as *data search* or *data discovery* practices, are commonly examined through user studies of particular data platforms and repositories (e.g. Borgman et al., 2019; Murillo, 2014; Wu, Psomopoulos, Jodha Khalsa, & de Waard, 2019). Zimmerman investigates data search practices directly, looking at the needs, discovery strategies, and criteria for evaluating data for reuse for a small group of environmental scientists. (2003, 2007, 2008). Recent work characterizes data search and evaluation practices across disciplinary domains (Gregory et al, 2019a; Gregory et al, 2019b) and by data professionals within and outside of academia (Koesten et al., 2017), relying primarily on in-depth interviews with data seekers or log analyses (Kacprzak, Koesten, Ibáñez, Simperl, & Tennison, 2017).

Surveys investigating data practices tend to use quantitative methods and focus on data sharing behaviors across disciplines (e.g. Tenopir et al., 2011; Tenopir et al., 2015; Kim & Zhang, 2015), within specific domains (Tenopir, Christian, Allard, & Borycz 2018; Schmidt, Gemeinholzer, & Treloar, 2016) or in different geographic locations (Ünal, Chowdhury, Kurbanoğlu, Boustany, & Walton, 2019). Publishers and data repositories conduct annual surveys tracing data sharing and management practices over time (e.g. Digital Science et al., 2018; Berghmans et al., 2017). Information about data search strategies, criteria important for reuse, and the role of social communications is found within surveys designed to develop data metrics (Kratz & Strasser, 2015) and to determine factors affecting data reuse (Kim & Yoon, 2017; Yoon, 2017).

Interest in designing tools specifically for data search is growing (Chapman et al., 2020), evidenced by the development of search engines for research data (Noy, Burgess, & Brickley, 2019; Scerri et al., 2017). Despite this trend, the limited amount of user interaction data restricts how these search tools are developed (Noy et al., 2019). There are also a growing number of policies regulating open data and data sharing (European Commission, 2019), which are seen as precursors to creating the ecosystem necessary for data discovery (Borgman, 2015b). These policies often do not accurately reflect the way that opening data and data sharing are enacted within communities (Noorman et al., 2018).

The sustainability and adoption of both search systems and data policies rely on understanding and building on extant practices (Schatzki, Knorr Cetina, & van Savigny, 2001). Our work aims to provide *evidence of practices of data seeking* and to inform the design of community-centric solutions and policies. To do this we take a broad approach, looking for commonalities which can be used for design, while also highlighting differences. We present a detailed examination of the behaviors of our respondents as they engage in discovering, evaluating and reusing data; this work also provides a starting point for future analyses.

## 3. Methodology

### 3.1 Survey design

We drew heavily on the findings of our earlier interviews with data seekers (Gregory et al, 2019a) and our analytical literature review (Gregory et al, 2019b) to design a survey addressing our principle research questions (see Table 1). Our research questions were informed by user-centered models of interactive information retrieval (e.g. Ingwersen, 1992; 1996; Belkin, 1993; 1996 ), particularly the synthesized model of an information journey proposed by (Blandford & Attfield, 2010; Adams & Blandford, 2005) which generally posits an actor/user with an (at times unrecognized) information need who engages in an iterative process of discovery, evaluation and use.

Our survey employed a branching design consisting of a maximum of 28 individual items; nine of these items were constructed to allow for multiple responses. In addition to write-in responses for expanding on "other" answers, the survey included three open-response questions. Respondents working as librarians or in research/data support also answered a slightly modified version of the survey.[1] Although we include some data from this group of "research support professionals" in the results presented here, most notably in Figure 10, the primary focus of this article is on researchers.

---

[1] The survey instrument is available in the supplementary material for this article.

| Research questions | Items |
|---|---|
| RQ 1: Who are the people seeking data? | Q1, L3, **L4**, **L5 D1**, D2, D3, D4, D5, D6, D7 |
| RQ 2: What data are needed for research and how are those data used? | *Q2*, **Q3**, **Q4**, Q5 |
| RQ 3: How do people discover data needed for research? | *Q5a*, Q6, Q7, **Q7a**, Q7b, Q8, **Q9**, Q10, *Q10a*, Q11, **Q11a** |
| RQ 4: How do people evaluate and make sense of data needed for research? | **Q9,** Q12, Q13, Q14, Q15 |

Table 1. Survey questions addressed by each research question. "L" questions were asked only to librarians/research support professionals; "D" questions were assigned to demographic section. Multiple response questions indicated in bold; open response questions in italics.

Four multiple response questions and their associated variables are of particular importance in our analysis; we include an overview of these variables to aid in navigating our results (Table 2).

| Item | D1 | | Q3 | Q4 | Q9 |
|---|---|---|---|---|---|
| **Variable** | **Disciplinary domains** | | **Types of data needed** | **Types of data uses** | **Social connections** |
| **Response options** | Agriculture | Engineering and Technology | Observational/empirical | Basis for a new study | Discover - Conversations with personal networks (e.g. colleagues, peers |
| | Arts and Humanities | Environmental Sciences | Experimental | Calibrate instruments or models | Discover - Contacting the data creator |
| | Astronomy | Health professions | Derived/compiled | Benchmarking | Discover - Developing new academic collaborations with data creators |
| | Biochemistry, Genetics, and Molecular Biology | Immunology and Microbiology | Simulated | Verify my own data | Discover - Attending conferences |
| | Biological Sciences | Materials Science | Other | Model, algorithm or system inputs | Discover - Disciplinary mailing lists or discussion forums |
| | Business, Management and Accounting | Mathematics | | Generate new ideas | Access - Conversations with personal networks (e.g. colleagues, peers |
| | Chemical Engineering | Medicine | | Teaching/training | Access - Contacting the data creator |
| | Chemistry | Multidisciplinary | | Prepare for a new project or proposal | Access - Developing new academic collaborations with data creators |
| | Computer Sciences / IT | Neuroscience | | Experiment with new methods and techniques (e.g. to develop data science skills) | Access - Attending conferences |
| | Decision Sciences | Nursing | | Identify trends or make predictions | Access - Disciplinary mailing lists or discussion forums |
| | Dentistry | Pharmacology, Toxicology and Pharmaceutics | | Compare multiple datasets to find commonalities or differences | Sensemaking - Conversations with personal networks (e.g. colleagues, peers |
| | Earth and Planetary Sciences | Physics | | Create summaries, visualizations, or analysis tools | Sensemaking - Contacting the data creator |
| | Economics, Econometrics and Finance | Psychology | | Integrate with other data to create a new dataset | Sensemaking - Developing new academic collaborations with data creators |
| | Energy | Social Science | | Other | Sensemaking - Attending conferences |
| | | Veterinary | | | Sensemaking - Disciplinary mailing lists or discussion forums |
| | | Information science | | | |
| | | Other | | | |

Table 2. Multiple response question numbers, variables, and response options of particular interest in our analysis.

The survey was scripted and administered with the Confirmit software (https://www.confirmit.com). We piloted the survey instrument in two phases. We scheduled appointments with four researchers, recruited via convenience sampling, and observed them as they completed the online survey. During these observations, we encouraged participants to think out-loud and ask questions. We used these comments to modify our questions before the next pilot phase. We then recruited an initial sample of 10,000 participants (using the recruitment methodology detailed below). Once one hundred participants had begun the survey, we measured the overall completion rate (41%), taking note of the points in the survey where people stopped completing questions. We used this information to further streamline the question order and to clarify the wording of some questions before recruiting our sample. A demographic analysis of the non-complete responses was not possible, as demographic information was collected at the end of the survey questionnaire.

## 3.2 Sampling and recruitment

Our population of interest consisted of individuals involved in research, in multiple disciplinary domains, who seek and use data they do not create themselves. This is a challenging population to target specifically, as information about who seeks and reuses data, particularly in a diversity of disciplines, is difficult to isolate (i.e. Yoon, 2014; Park, You & Wolfram, 2018). We therefore recruited our sample from a very broad population of researchers, hoping to capture a diverse sample of data seekers.

We sent recruitment emails to a random sample of an additional 150,000 authors who are indexed in Elsevier's Scopus database and who have published in the past three years. The recruitment sample was constructed to mirror the distribution of published authors by country within Scopus. Recruitment emails were sent in two batches, one of 100,000 and the other of 50,000, two weeks after the first batch. One reminder email was sent to encourage participation. A member of the Elsevier Research and Academic Relations team created the sample and sent the recruitment letter, as access to the authors' email addresses was restricted. Potential participants were informed that the purpose of the survey was to investigate data discovery practices. We therefore assume that participants who completed the survey are in fact data seekers.

We received 1637 complete responses during a four-week survey period in September-October 2018. We recruited an additional 40 participants by posting to discussion lists in the library and research data management communities, for a total sample of 1677 complete responses, yielding a response rate of 1.1%. This response rate is calculated using the total number of recruitment emails distributed. It is likely that not all 150,000 individuals receiving recruitment emails match our desired population of individuals who search for and reuse data.

Of the recruited participants, 2,306 individuals clicked on the link to the survey, but did not complete it. Forty-two percent of people who engaged with the survey completed the questionnaire, similar to the completion rate in our pilot phase. Of the individuals who did not complete the survey, fifty percent viewed the introduction page with the informed consent statement, but did not click through to the survey itself. This could be because participants were not interested in the scope of the survey, had negative feelings about the funders or institutions involved, or that they did not agree with the information presented in the consent form. Seventy-five percent of individuals who did not complete the survey stopped responding by the fifth question, which was still within the first of the four sections of the survey.

3.3 Analysis

We used the statistical program R to perform our analysis, in particular the Multiple Response Categorical Variable (MRCV) package (Koziol & Builder, 2014a; Koziol & Builder, 2014b) to analyze questions with multiple possible responses. For these questions, we tested for multiple marginal independence (MMI) or simultaneous pairwise marginal independence (SPMI) between variables using the Bonferroni correction method. This method calculates a Bonferroni-adjusted p-value (Dunn, 1961) for each possible 2x2 contingency table that can be constructed from a question's variables. These individually adjusted p-values are then used to create an overall adjusted p-value which indicates if MMI or SPMI exist; specific associations between variables can then be identified by comparing the p-values from the individual 2 x 2 tables to the set significance level ($\alpha$ = 0.05). (Bilder, & Loughin, 2015). This method was used to detect the correlations presented in Figure 6, Figure 8, Figure 15 and Table 4. This test can produce overly conservative results, particularly when analyzing questions with many variables. Nonetheless, this approach is preferred to traditional tests for independence, as it takes into consideration the fact that a single individual can contribute to multiple counts within a contingency table. We coded and analyzed open response questions in NVivo using a combined deductive and general inductive approach to thematic analysis (Thomas, 2006) and used R, Gephi and Tableau to create plots and visualizations.

3.4 Reporting

Significant associations between variables are often reported by listing p-values in tabular format for each combination of possible responses. Due to the complexity of our questionnaire, in particular the number of multiple response variables (Table 2), we present significant associations using visualizations. These visualizations indicate if an association between variables exists; they do not indicate the value of the p-values themselves. We do this in order to increase the understandability of our results and to make them usable for a wider audience. Tables with the adjusted p-values are included in the supplementary material for this article.

3.5 Limitations

The sampling methods, the survey design, and our analytical methods have both strengths and limitations (see Box 1). Our data and analysis are descriptive, not predictive, and only represent the practices of our respondents - a group of data-aware people already active in data sharing and reuse and confident in their ability to respond to an English-language survey. We also have limited knowledge about individuals who did not respond to the survey. The analysis presented here is not generalizable to broader populations, but rather depicts the behaviors and attitudes of the approximately 1700 respondents to our survey.

| **Strengths and limitations of this study** |
|---|
| **Sample** |
| • Responses are from globally-distributed, multi-disciplinary individuals involved in research |

> - Number of responses (n=1677)
> - Possible sampling bias due to coverage of Scopus
> - Response rate of 1.1%. Limited information about non-responders.
>
> **Survey questionnaire**
> - Multiple response and open response questions elicited rich qualitative and quantitative data
> - Survey questions based on past empirical work
> - Responses represent self-reported behaviours and attitudes
> - Responses to multiple choice questions shaped by the options provided. Open response option included for each multiple choice question.
>
> **Analysis**
> - Analysis represents discovery phase for a rich, hard-to-obtain dataset. Exploratory quantitative results combined with qualitative analysis.
> - Large sample size especially important in qualitative analysis.
> - Provides directions for future areas of more detailed and statistically powerful quantitative analysis
> - Exploratory tests for statistical significance are based on pairwise correlations.
> - Statistically significant results may be impacted by possible bias in the sample.
> - Complete-case analysis. Partial responses were not included.

Box 1. Strengths and limitations of sampling, questionnaire design and analysis methods

The majority of our sample consists of researchers who have published in a journal indexed in the Scopus database. Certain disciplinary domains are better represented within Scopus; most notably the arts and humanities are not as well covered (Mongeon & Paul-Hus, 2016; Vera-Baceta, Thelwall, & Kousha, 2019). Scopus has an extensive and well-delineated review process for journal inclusion; as of January 2020, 30.4% of titles in Scopus are from the health sciences; 15.4% from the life sciences; 28% from the physical sciences and 26.2% from the social sciences (Elsevier, 2020). While the policies and content of Scopus could produce a potential bias in our sample, drawing from this population also helped us to target our desired population of researchers across domains.

Self-reported responses could be impacted by respondents' desire to give socially acceptable answers. Respondents may also have interpreted the survey questions in different ways; responses could be influenced by English-language fluency, individual interpretations of the wording of questions, the multiple choice options and Likert scales provided, and the ordering of the questions themselves. We attempted to counter these issues by implementing a two-stage pilot phase and by providing open response options for multiple choice questions.

Our findings are based only on complete survey responses. While this facilitated our analysis, it is possible that this introduced additional biases to our sample as well as reducing potential statistical power (as described in Pigott, 2001). The test for independence that we apply to identify correlations among questions with multiple response variables has been suggested to err on the conservative side, (Bilder, 2015), making it possible that some correlations were not detected with this method. The Bonferroni adjustment is an approximation; known issues include the identification of false negatives, particularly when comparing large numbers of variables, and determining the number of comparisons to use in the calculation (McDonald, 2014). For these reasons, we present our quantitative results as an initial exploration of the data, to guide deeper future statistical analyses.

3.6 Ethics and Data Availability

We received ethical approval from Maastricht University for the study. Participants had the opportunity to review the informed consent form prior to beginning the survey and indicated their consent by proceeding to the first page of questions.

The data from this survey are available in the DANS-EASY data repository under a CC-BY-4.0 license (Gregory, 2020).

**4. Results and Analysis**

We present our results according to the research questions presented in Table 1. Each subheading provides an answer to the proceeding research question, which is then further discussed and supported by the survey data. The research questions are therefore answered in the course of the results section, before the primary takeaways are summarized in the Conclusion. We first examine characteristics of the data seekers responding to our survey, and then proceed to look at their data needs, search and discovery strategies, and evaluation and sensemaking practices.

4.1 RQ 1: Who are the people seeking data?

***Respondents are globally distributed and have research experience***
Respondents employed in 105 countries completed the survey. The United States, Italy, Brazil and the United Kingdom were among the most represented countries (Figure 1a). This does not directly correspond with the sampling distribution (Figure 1b), where the largest difference between recruited participants and respondents occurred in China. This lower response rate could be due to language differences, perceived power differences (Harzing, 2006), or a lack of tradition in responding to research requests from researchers from other countries (Wang & Saunders, 2012). It could also indicate that data seeking is not a common practice.

The majority of survey respondents were researchers (82%) and worked in universities (69%); 40% of respondents have been professionally active for 6-15 years (Table 3). With the exception of participants recruited specifically from the library and research data management communities (or "research support professionals"), our recruitment methodology ensures that all respondents are published authors, making it likely that they have been involved in conducting research in either their past or current roles. Nearly half of those working in research support also need secondary data for their own research, in addition to supporting researchers or students.

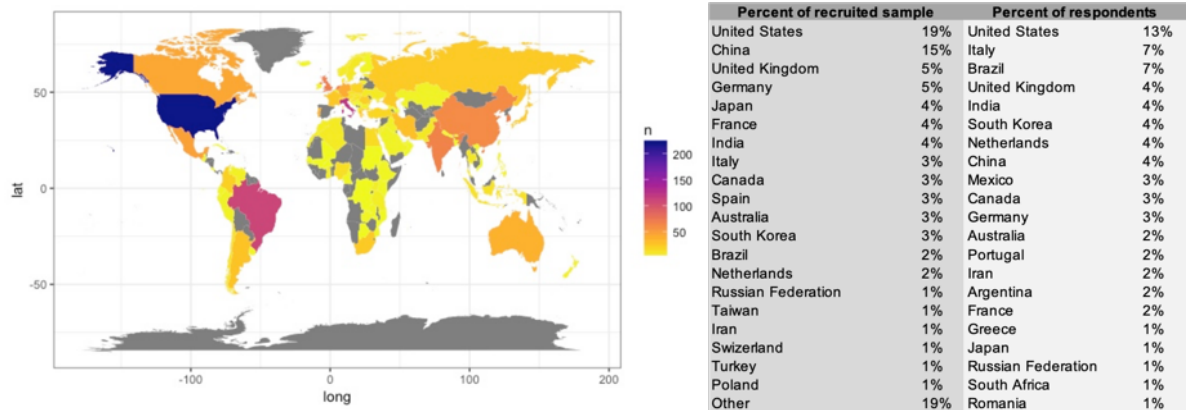

Figure 1. a) Number of respondents by country of employment (n=1677). b) Percent of recruited participants by country compared to percent of respondents by country (n=1677)

| Role | Number | Percent |
|---|---|---|
| Researcher | 1372 | 82 |
| Student | 73 | 4 |
| Manager | 54 | 3 |
| Research support | 47 | 3 |
| Other | 131 | 8 |
| **Organization** | | |
| University | 1150 | 69 |
| Research institution | 287 | 17 |
| Government agency | 74 | 4 |
| Corporate | 68 | 4 |
| Independant archive | 2 | 0 |
| Other | 96 | 6 |
| **Professional experience** | | |
| 0-5 | 206 | 12 |
| 6-15 | 677 | 40 |
| 16-30 | 502 | 30 |
| 31+ | 292 | 17 |

Table 3. Role, place of employment and years of professional experience of respondents (n=1677).

*Respondents support data sharing and reuse*

While eighty percent of all respondents reported sharing their own data in the past, participants with longer careers have done so slightly more often. Eighty-nine percent of respondents who have worked for 31+ years reported having shared their data, compared to 77% percent of respondents working for less than five years. Personal attitudes towards data sharing and reuse differ from the perceived attitudes of peers, disciplinary communities and institutions (Figure 2). The majority of survey respondents are proponents of sharing their research data; they believe that the other groups identified in Figure 2 are less supportive of data sharing. This also points to a possible bias in our data, suggesting that people who have not shared their own data or who do not support data sharing were less likely to respond to our survey. Alternately, respondents could have felt compelled to provide a socially desirable answer, feeling a positive response was more acceptable. Respondents indicated that they believe data sharing is more strongly supported by their direct co-workers than by their disciplinary communities or institutions; they are most uncertain about the attitudes of their institutions. A similar pattern exists for attitudes toward data reuse, although there is more uncertainty involved.

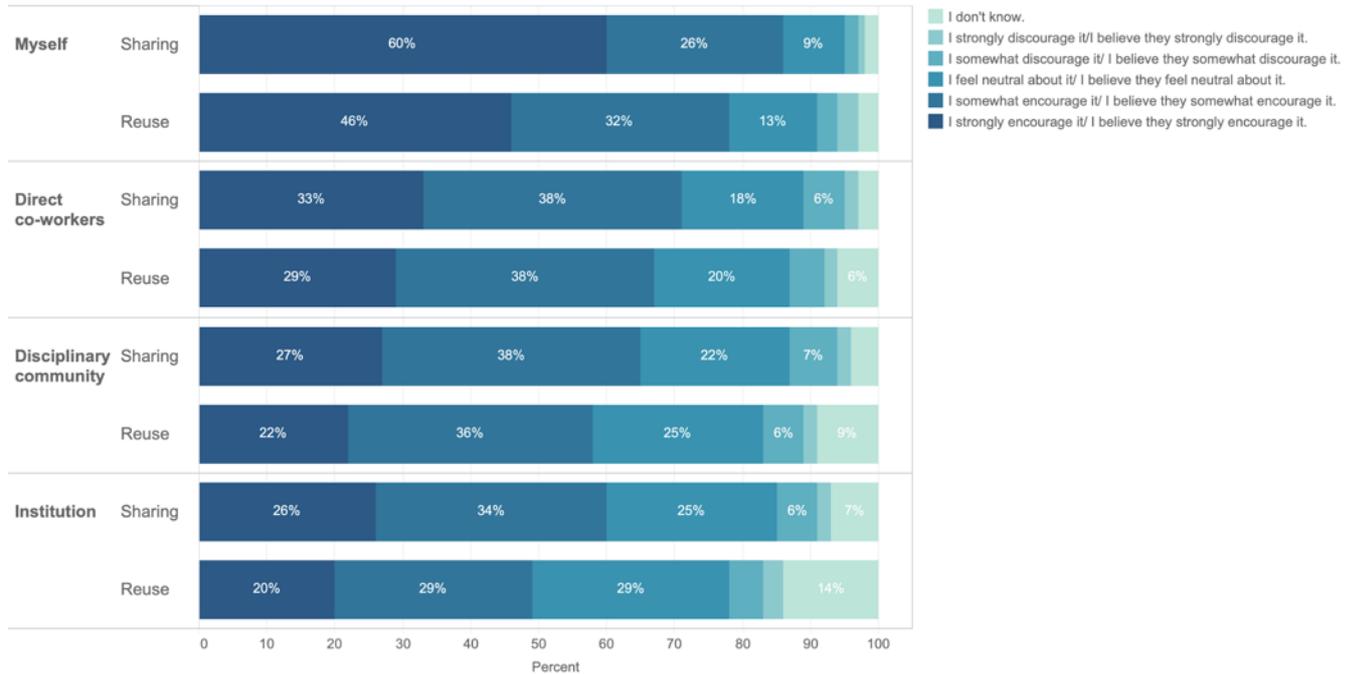

Figure 2. Respondents' beliefs about how they and other groups feel about data sharing and reuse (n=1677). Percent denotes percentage of respondents.

*Respondents belong to multiple, overlapping domains*

Respondents indicated their domain(s) of specialization from a list of 31 possibilities. This disciplinary list was originally used in a survey measuring data sharing practices across disciplines (Berghmans et al., 2017). Engineering and technology was selected most often, followed by biological, environmental, and social sciences (Figure 3). Approximately half of the respondents selected two or more domains, with one quarter selecting more than three. This could be a factor of varying levels of granularity of the disciplinary list; it could also indicate that participants found it challenging to choose a single domain that captures the complexity of their expertise.

Figure 3 depicts the disciplinary overlaps among respondents, showing which domains were selected in conjunction with each other. The figure reveals expected disciplinary overlaps, e.g. between information science and computer science, between medicine and health professions and between material science and chemistry. Other overlaps are perhaps less expected, for example between social science and engineering and technology or between arts and humanities and computer science, which could be indicative of the use of digital humanities methodologies among our participants. Seventy-one percent of respondents who selected engineering and technology chose at least one other discipline, most frequently computer science, environmental science, material science, and energy, as is visible in Figure 3.

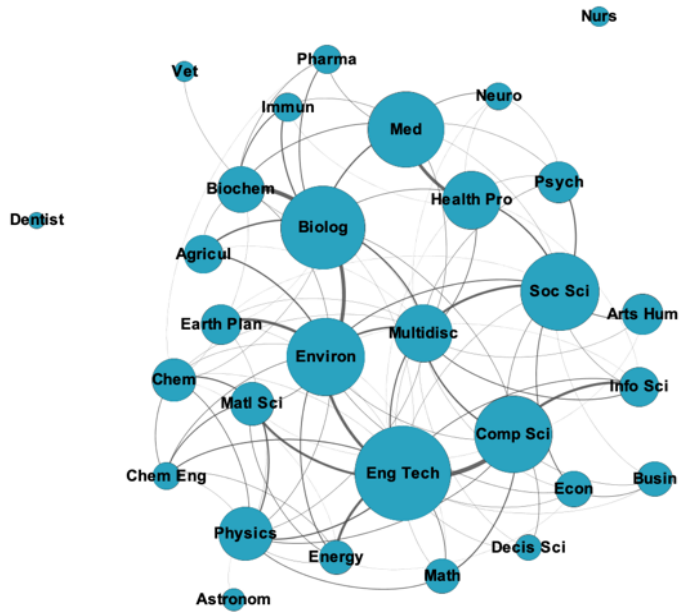

Figure 3. Disciplinary domain responses (n=3431). Node size represents the number of responses from a discipline. Width of the edges represents the number of times two disciplines were selected in conjunction with each other. Only edges with a weight greater than 20 are shown; some edges pass behind other nodes. Graph created in Gephi; graph shape set using ForceAtlas algorithm, repulsion strength=9000; other parameters set to default.

We also identified individuals who selected only one discipline. The greatest number of single-discipline responses were in the domains of medicine, social science, engineering and technology and computer science (Figure 4). In future sections of this paper, we augment our analysis of respondents across all disciplinary domains with an occasional analysis of a subset of the individuals selecting only one discipline.

In this disciplinary subset, we included domains with more than 40 respondents, with the exception of the "other" category, as well as other domains whose data practices have been well-documented in the literature, i.e. astronomy, environmental sciences and earth and planetary sciences (see Gregory et al, 2019b for a review of the literature exploring these disciplines). Disciplines which are included in this disciplinary subset are marked with a double asterisk in Figure 4.

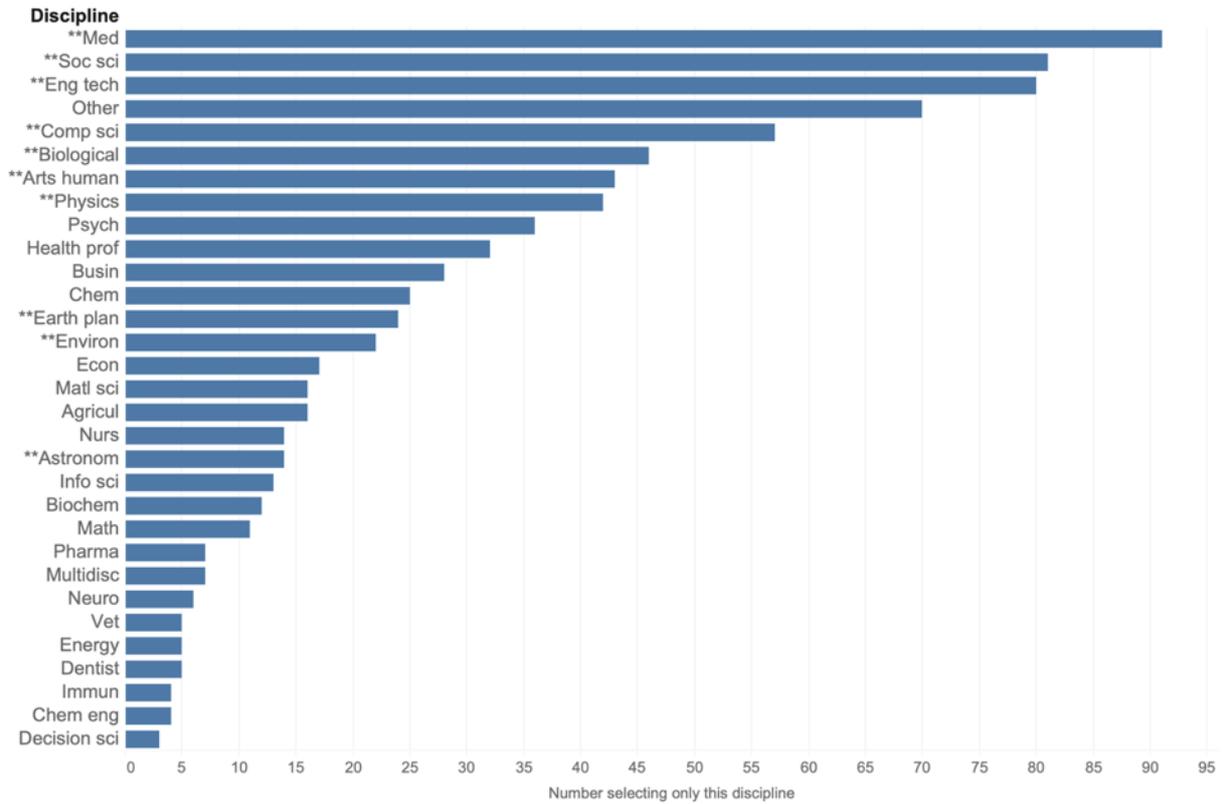

Figure 4. Number of respondents who selected only one discipline (n=836). Disciplines included in the subset for further analysis marked with **.

4.2 RQ 2: What data are needed for research and how are those data used?

***Data needs are diverse and difficult to pigeonhole.***

To provide a high-level view of their data needs, respondents selected the type(s) of data that they need in their work from a list derived from the categories of research data proposed by the United States National Science Board (2005) and used in (Berghmans et al., 2017; Gregory et al, 2019b). While observational/empirical data were selected most frequently, 50% of participants also indicated that they need more than one type of data. Figure 5 represents the number of respondents selecting individual and multiple data types.

Please select the options that describe the secondary data that you (might) need.

☐ **Observational or empirical** (e.g. sensor data, survey data, interview transcripts, sample data, neuroimages, ethnographic data, diaries)

☐ **Experimental** (e.g. gene sequences, chromatograms, toroid magnetic field data)

☐ **Simulation** (e.g. climate models, economic models)

☐ **Derived or compiled** (e.g. text and data mining, compiled database, 3D models)

☐ **Other,** Please specify __________

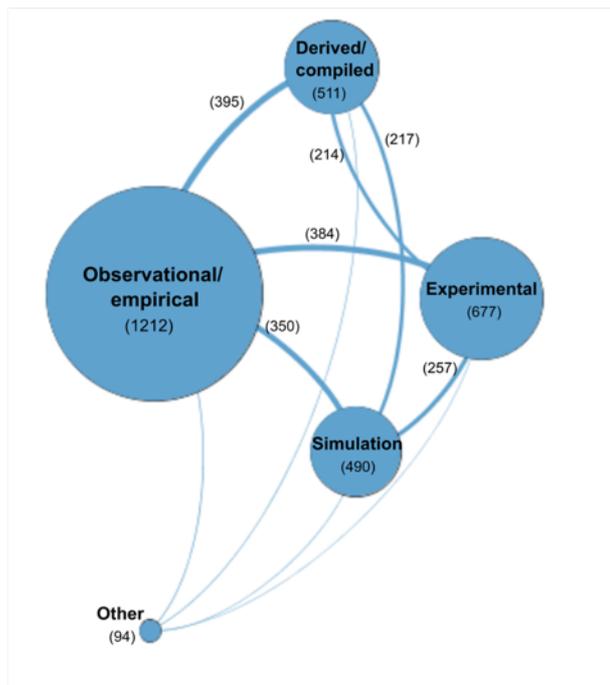

Figure 5. Question from survey with descriptions of data types. Node size in visualization represents number of responses for each data type (n=2984). Edges represent number of times both data types were selected in conjunction with each other.

We asked participants to further expand on their data needs in an open response question, which seventy-eight percent of respondents completed. We compared these responses to the types of data that participants selected, paying special attention to those who chose multiple data categories. Respondents selecting multiple data types appear to do so for different reasons. Some require a variety of topic-specific data to conduct their research; others use a variety of data, regardless of topic, as long as it matches format and structure requirements. The following participant indicated needing observational, experimental and derived data.

> [I need] large and small datasets that students can use for Data Science skill development. For example, transactional data from business, medical data such as interactions between patients and doctors, descriptive medical histories. Data need to be complex enough to be interesting and able to be parsed. (Respondent ID 613).

Data are difficult to categorize (Borgman, 2015b; Leonelli, 2019); in part because people may define data categories differently. While the majority of respondents stating they use census data selected observational data, others did not, choosing instead derived/compiled data. While the majority of individuals using literature corpora selected derived/compiled data, a minority selected "other," apparently not knowing which category best fit their data.

Associations (detected using the test for independence described in the Methodology section above) between disciplinary domain and the type of data needed are shown in Figure 6. The highest number of disciplinary associations were detected for experimental data; arts and humanities was associated with the greatest number of data types, although surprisingly not with observational data. There are also domains where no associations were detected. This could be due to the composition of our sample; it could also be taken as evidence for the diversity of data types that participants within disciplines need.

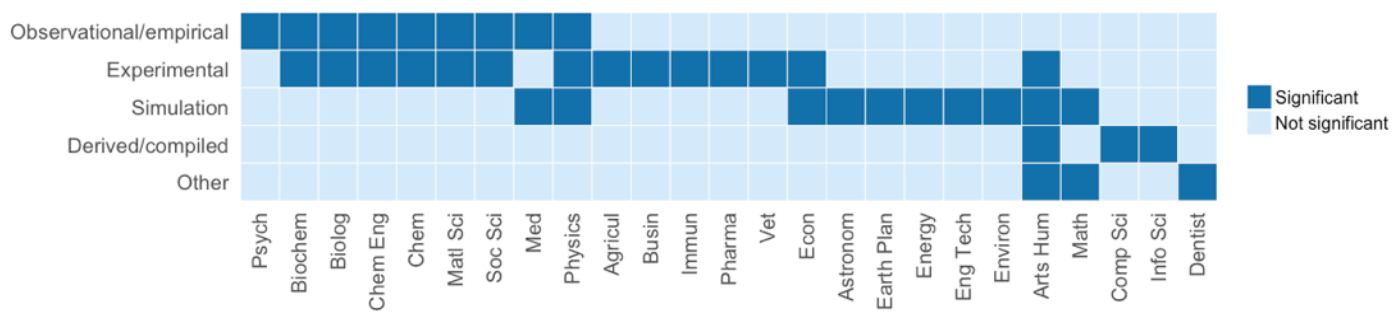

Figure 6. Associations between disciplinary domain and needed data. Associations detected using adjusted Bonferroni test for simultaneous pairwise marginal independence (n=1677; significance level: $p < 0.05$).

Slightly more than half of respondents reported needing data outside of their disciplines. This pattern holds across domains, with the exception of environmental science, where more individuals need data external to their domain (65%), and medicine, where the trend was reversed, with less than half needing this type of data (43%).

*Data uses are diverse but limited to a core set*

Although using data in ways that support research is well-documented (e.g. Wallis, Rolando, & Borgman, 2013), using data to drive new research or in teaching is not as well documented (Gregory et al, 2019b). The majority of respondents (71%) selected using data as the basis for a new study, which is in line with other very recent research (Pasquetto, Borgman, & Wofford, 2019); half of respondents selected needing data to use in teaching (Figure 7). Other uses include experimenting with new methodologies and techniques such as developing data science skills or completing particular data-related tasks, i.e. trend identification or creating data summaries (as suggested by Koesten et al., 2017). Less than two percent of respondents indicated needing data for other purposes, suggesting that the list of uses in our survey is fairly complete.

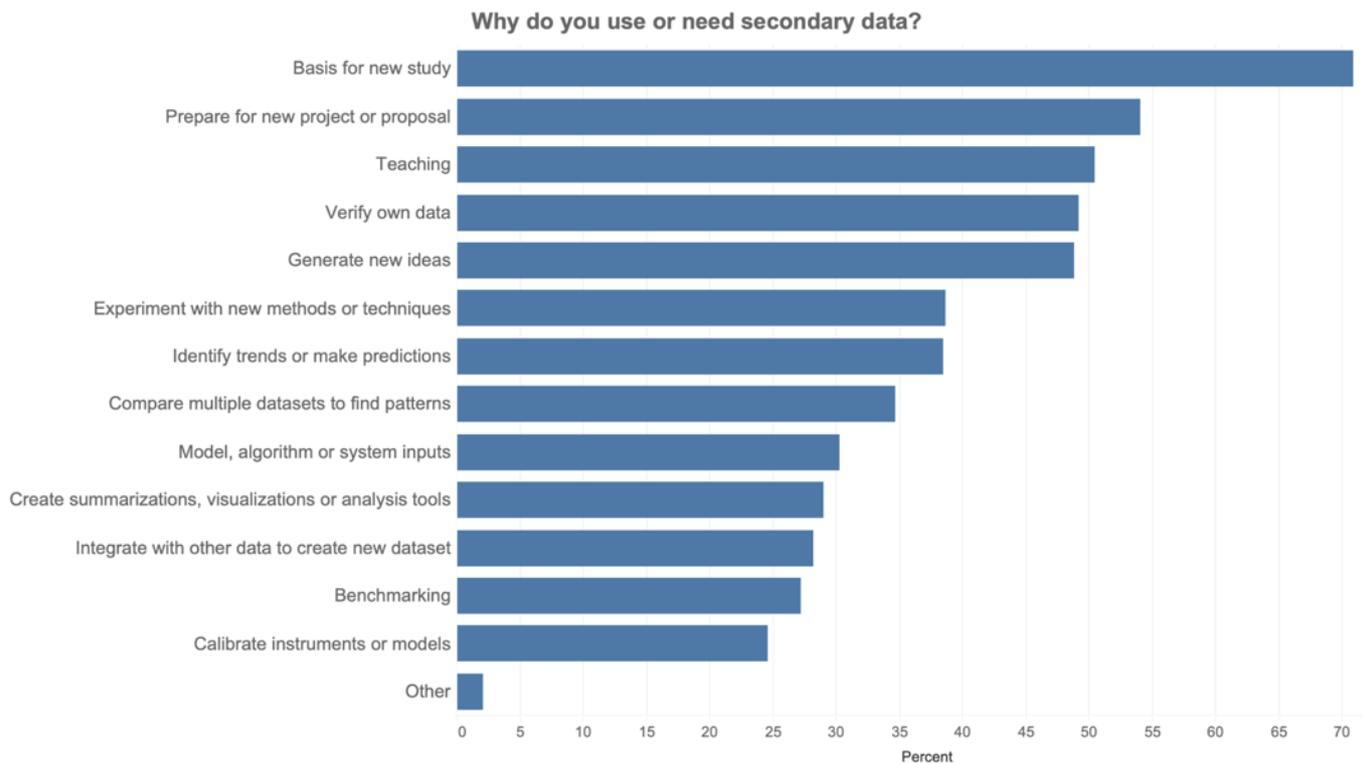

Figure 7. Reasons why respondents need secondary data. Multiple responses possible. Percent denotes percentage of respondents (n=1677).

*Data uses span research phases*

Table 4 presents significant associations between the data uses shown in Figure 7 and the types of needed data shown in Figure 5. We also tested to see if significant associations exist between the different types of data uses (*Table 4: Associated with these types of data uses)*. We recognize that associations between uses are difficult to interpret; if an individual selected teaching, i.e., it may or may not be related to their selection of another data use. To examine these associations through a different perspective, we organize them along a conceptual typology, which we propose here, based on different phases of research.

To create this typology, we drew on the associations we identified, existing literature, and extant models of research lifecycles. In Table 4, we agree with van de Sandt, Dallmeier-Tiessen, Lavasa, & Petras (2019) that reuse is one type of data use, but similar to Fear (2013), we view reuse as referring to using data as the basis for a new study. Our typology has similarities with models of the research lifecycle process (e.g. Jisc, 2013), but it also differs from these models. Typical models of research lifecycles portray research that uses data created by the researcher, rather than secondary data. They also tend to reduce the complexity of research processes (Borgman, 2019), ignore the interwoven nature of tasks involved in research (Cox & Wan Ting Tam, 2018) and depict data cycles as independent workflows (e.g. UK Data Service, 2019). With this typology, we attempt to nuance the uses of secondary data throughout phases of research, recognizing how data uses are associated with multiple data types and uses and highlighting that these uses are associated with multiple work phases.

Depending on research practices, the data uses in our typology could occur in different phases. Uses that could particularly fall into two bordering categories are marked in grey. Integrating different datasets could occur when conducting research, e.g., or verification of data could also be considered to be part of data analysis. Although not marked in grey, data analysis and sense-making tasks are likely to occur throughout all phases of research. This is reflected in the results, where analysis activities are associated with every other data use, with the exception of instrument or model calibration.

| Research/work phase | Types of data uses | Associated with these types of needed data | Associated with these types of data uses |
|---|---|---|---|
| Reuse | Basis for new study (new study) | observational/empirical | new project, new ideas, integration, comparison, trends, teaching |
| Project creation and preparation | Prepare for new project or proposal (new project) | | new study, new ideas, integration, new methods, verification, trends, comparison, summaries/vis/tools, teaching |
| | Generate new ideas (new ideas) | | new study, new project, integration, new methods, verification, trends, comparison, summaries/vis/tools, teaching |
| | Integrate with other data to create new dataset (integration) | observational/empirical, derived/compiled | new study, new project, new ideas, new methods, inputs, trends, comparison, summaries/vis/tools |
| Conducting research | Experiment with new methods or techniques - e.g. data science skills (new methods) | experimental, simulation, derived/compiled | new project, new ideas, integration, inputs, calibration, benchmarking, verification, trends, comparison, summaries/vis/tools |
| | Model, algorithm or system inputs (inputs) | simulation, derived/compiled | integration, new methods, calibration, benchmarking, trends, comparison summaries/vis/tools |
| | Calibrate instruments or models (calibration) | experimental, simulation, derived/compiled | new methods, benchmarking, inputs, verification |
| | Benchmarking (benchmarking) | simulation, derived/compiled | new methods, inputs, calibration, verification, trends, summaries/vis/tools |
| | Verify own data (verification) | experimental, simulation, derived/compiled | new project, new ideas, new methods, calibration, benchmarking, comparison, teaching |
| Data analysis and sensemaking | Identify trends and make predictions (trends) | observational/empirical, simulation, derived/compiled | new study, new project, new ideas, integration, new methods, inputs, benchmarking, comparison, summaries/vis/tools |
| | Compare multiple datasets to find commonalities or differences (comparison) | experimental, derived/compiled | new study, new project, new ideas, integration, new methods, inputs, verification, trends, summaries/vis/tools |
| | Create summaries, visualizations or analysis tools (summaries/vis/tools) | observational/empirical, simulation derived/compiled | new project, new ideas, integration, new methods, inputs, benchmarking, trends, comparison, teaching |
| Teaching | Teaching/training (teaching) | observational/empirical | new study, new ideas, new project, verification, summaries/vis/tools |

Table 4. Associations between types of data use, needed data type and other data uses. Grey areas represent uses that could fall within multiple bordering research phases. Colors correspond to research/work phases. Associations detected using adjusted Bonferroni test for simultaneous pairwise marginal independence (significance level: $p < 0.05$).

Using data as the basis for a new study is associated in our results with needing observational data. Among our respondents, observational data are not significantly associated with the tasks in the "conducting research" phase, but they are associated with activities in both data analysis and with teaching. Experimental data are related to uses involved in conducting research, as well as to comparison. Derived data are associated with all activities in both the research process and data analysis/sensemaking phases; simulation data are primarily associated with conducting research.

Using data as the basis for a new study is associated with uses such as project creation and preparation, data analysis and sense-making, or teaching, but is not associated with any uses in the conducting research phase. Teaching is associated with only a few other uses, most of which fall within the project creation phase. Data integration is associated with activities across phases of research; calibration, however, is exclusively associated with other research process tasks.

*Data uses are common, but their enactments are complex*
Disciplinary domains also shape how data are used. Figure 8 indicates significant associations between disciplines and uses in our sample. Most of these associations are for data uses that fall within the conducting research phase of the above typology, particularly using data as inputs and for calibration, and domains that typically make use of modelling or computational research methods. Due to the large number of variables compared in Figure 8 and the limitations of our test for significance, these associations in particular represent initial results.

To complement the associations presented in Figure 8, we also looked at the subset of our data for individuals selecting only one discipline (indicated in Figure 4). Within these disciplinary groups, we also found that respondents use data for a variety of purposes, rather than for just one type of use. While researchers in multiple domains may use data for the same purposes, uses will be enacted in different ways and have different meanings in various disciplines and contexts (Borgman, 2015b, Leonelli, 2015). While 39% of computer scientists in the disciplinary subset and 35% of those in the arts and humanities selected using data for verification, for example, the actual practice and the meaning of verifying data will be different in each of these disciplines.

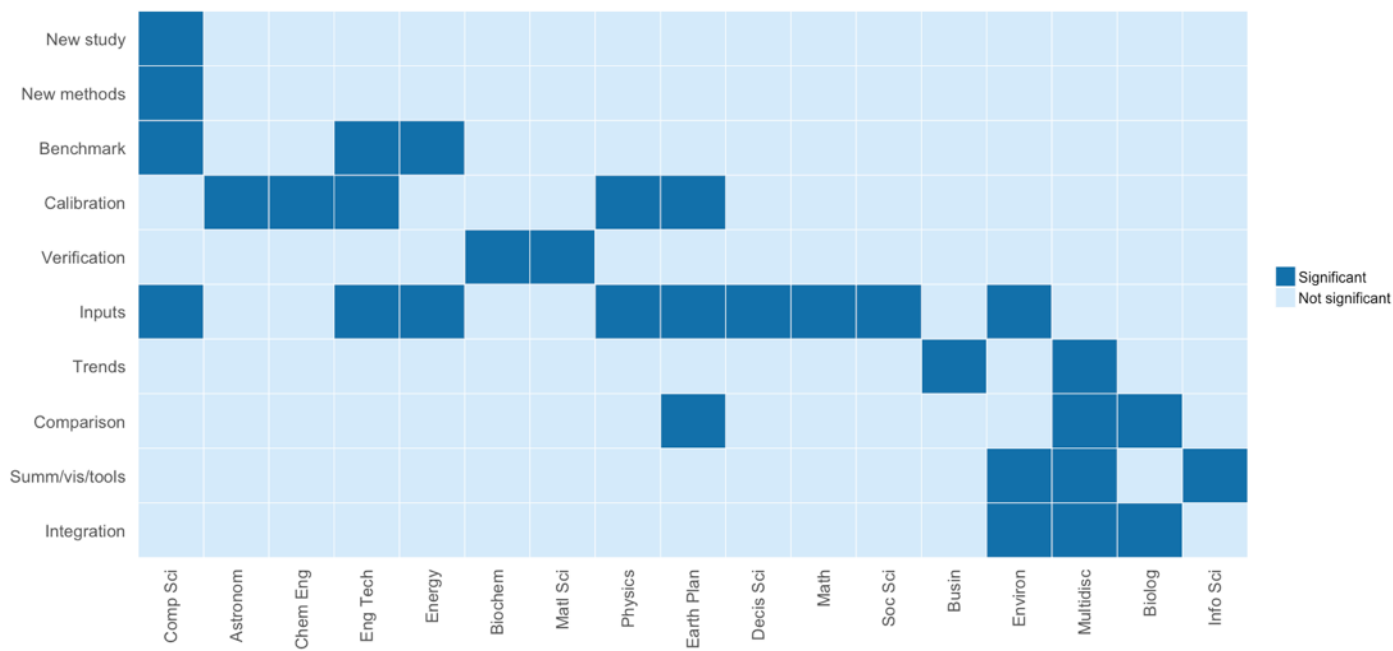

Figure 8. Associations between disciplinary domain and data use. Associations detected using adjusted Bonferroni test for simultaneous pairwise marginal independence (n=1677; significance level: p < 0.05).

4.3 RQ 3: How are people discovering data?

Researchers believe that discovering data is a sometimes challenging (73%) or even difficult (19%) task. The greatest challenge researchers face is that data are not accessible, i.e. data are not available for download or analysis, (68% of researcher respondents), followed by the fact that data are distributed across multiple locations (49%). One third of these respondents identified inadequate search tools, a lack of skill in searching for data, or the fact that their needed data are not digital as being additional challenges.

*Via academic literature*

Thirty percent of respondents reported no difference in how they find literature and how they find data. Figure 9 presents the disciplines selected by these respondents. Some of these respondents chose multiple disciplines; the analysis in Figure 9 is therefore not limited to respondents in the disciplinary subset. Disciplines with data repositories that are closely linked with systems for searching the academic literature of that field, such as in the biomedical sciences and physics (i.e. the resources from the National Library of Medicine[2] and HEP-INSPIRE[3] database, respectively) rank more highly. Respondents in disciplines where data repositories and academic literature databases are traditionally less integrated with each other, i.e. in the social sciences (e.g. the ICPSR[4] database) or the environmental sciences (e.g. PANGAEA[5]), tend to have more distinct discovery practices.

---

[2] https://www.ncbi.nlm.nih.gov/guide/all/
[3] http://inspirehep.net

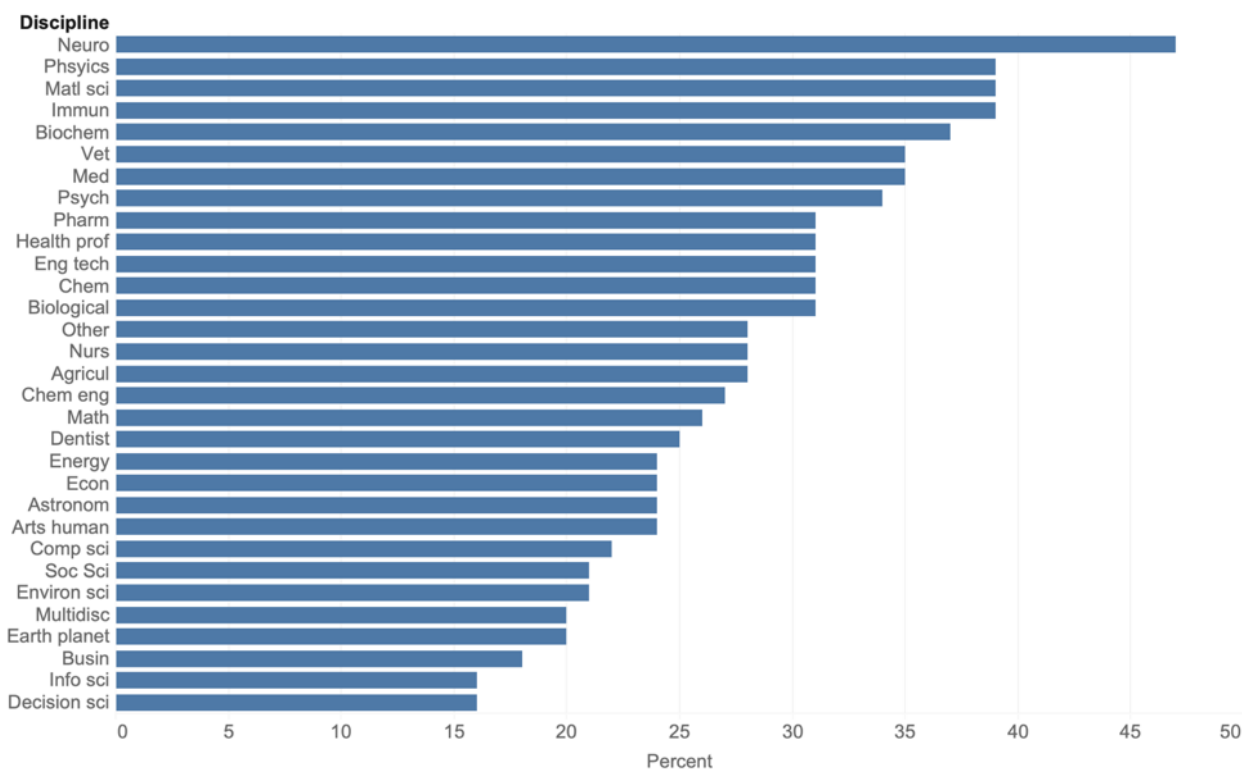

Figure 9. Percentage of responses (n=959) by discipline for individuals reporting no difference in literature and data search strategies. Multiple responses possible.

Fifty-two percent of respondents stated that their processes for finding literature and data are sometimes the same and are sometimes different; whereas they were always different for 18%. Respondents saying that the two processes are sometimes or always different (n= 1178) were asked to explain the differences in an open response question.

One of the key differences participants identified between literature search and data discovery are the sources that are used.

> I check other channels for data than for literature, e.g. if a project produces data, I check the project's site directly for their data and hope for links to repositories. (Respondent ID 4001)

> Academic literature could be found through different portals...To receive data, one often needs to know where to find it. For example, the name on the database and then contact the administrator for the database if you can't extract the data directly from the database. (Respondent ID 4008)

---

[4] https://www.icpsr.umich.edu/icpsrweb/
[5] https://www.pangaea.de

Yet the academic literature itself is a key source for discovering data for researchers, as are general search engines (e.g. Google) and disciplinary data repositories (Figure 8). Research support professionals rely less on the literature, more frequently turning to a variety of sources in their search for data (Figure 10). The importance of literature, search engines and domain repositories across disciplines supports the findings from earlier work (Kratz and Strasser, 2015).

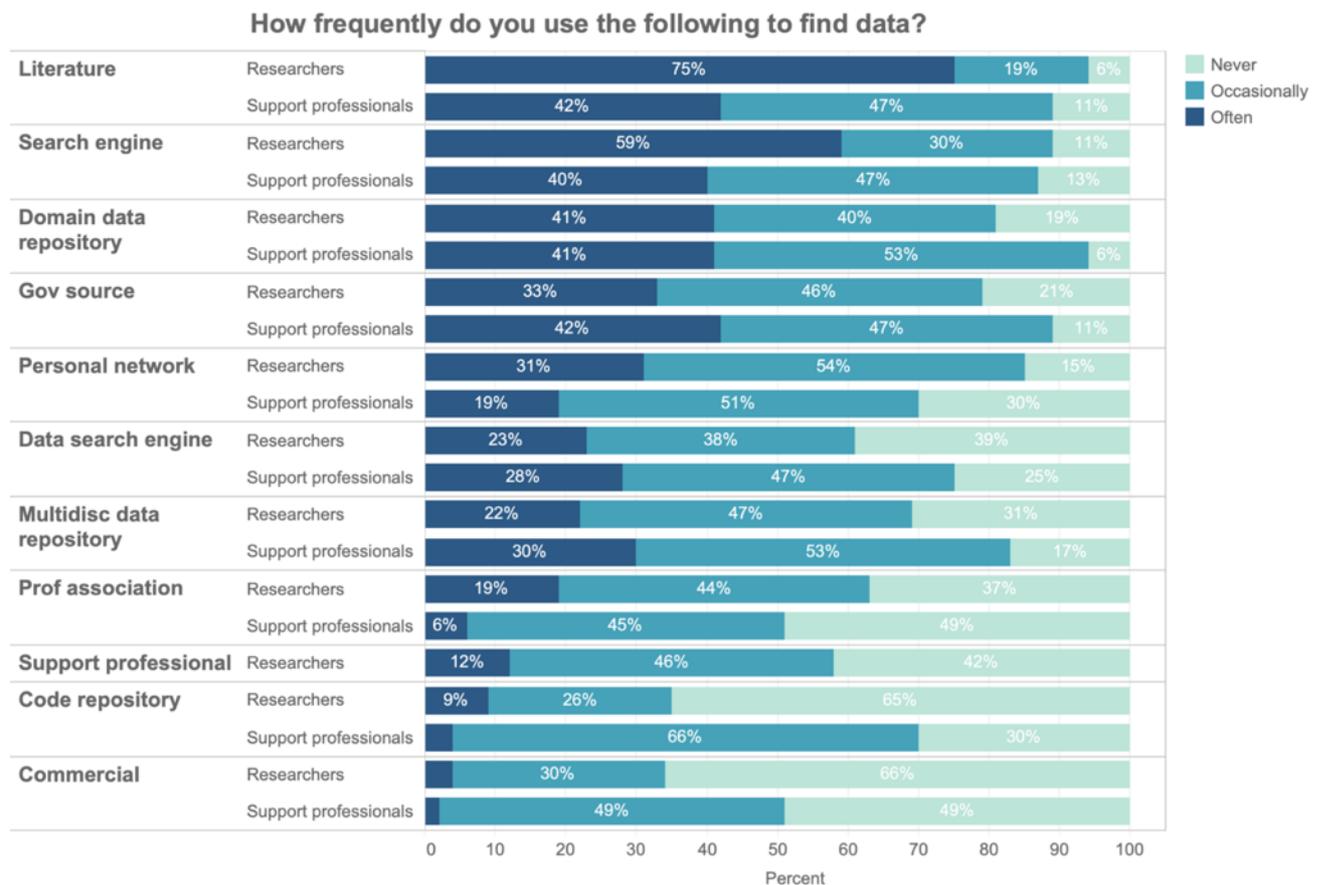

Figure 10. Sources used to find data by researchers (including students, managers, and others, n = 1630) and research support professionals (n=47). Percent denotes percentage of respondents for each category. Listed in order of decreasing importance for researchers.

The distribution of sources presented in Figure 10 generally holds across disciplines for researchers; literature, followed by search engines or domain repositories, are reported as being used most often in nearly all domains. There are some disciplinary differences, identified by looking at the subset of respondents selecting only one discipline. In the arts and humanities, e.g., turning to research support professionals was selected more often than in other disciplines; in computer science, 76% of respondents occasionally or often consult code repositories. (A breakdown of the use of sources according to domain for our disciplinary subset is included in the supplementary materials).

Respondents use literature as a source of data - plucking data from reported tables and graphs[6] - but they also use the literature to track down the original data, making use of behaviors common in literature searching, such as citation chasing (Figure 11; the distribution presented in Figure 11 remains the same when looking at the percentage of overall responses to this question.) It is common for respondents to first find the literature, and then use the literature as a gateway to locating the data. This strategy is often planned, but it also happens serendipitously while reading or searching for literature. Roughly two thirds of participants also often or occasionally find data serendipitously outside of the literature (e.g. via email or conversations with colleagues) or in the course of sharing or managing data.

> Finding data is different because it often occurs as a result of finding academic literature. (Respondent ID 738)

> Literature is more direct; data is more like "bonus" finds. One finds interesting data in other contexts of work in a publication, one can contact the author to ask for the data. (Respondent ID 3179)

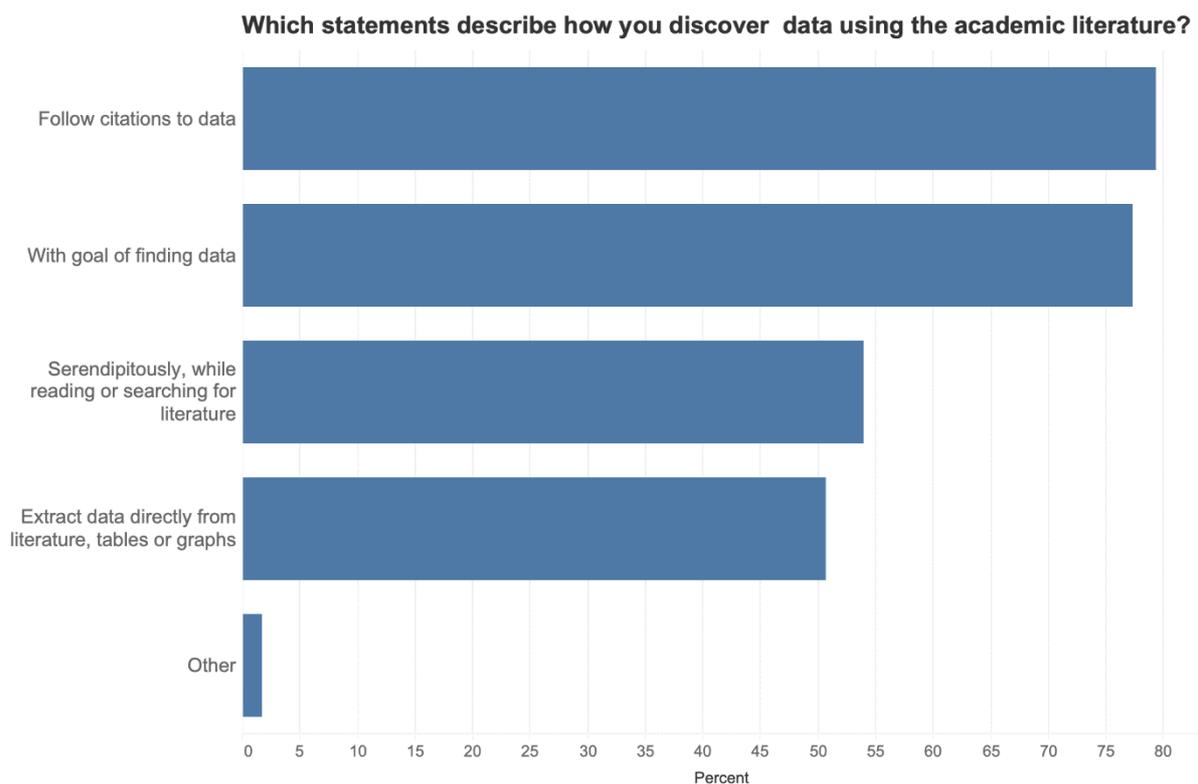

Figure 11. Strategies for using the academic literature to discover data. Question asked to respondents who indicated using literature as a source. Percent denotes percentage of respondents; multiple answers were possible (n=1573).

### *Via social connections*

---

[6] This practice in particular confirms the importance of asking users what they actually do, rather than assuming an ideal notion of data search and reuse.

Using social connections and personal outreach to discover and access data is another important difference identified between literature search and data search. This is reflected in Figure 10 where only 15% of researchers never make use of personal networks in data discovery.

> Unlike academic literature where you get the data by accessing the journal, finding data often requires contacting the institution that created the data. (Respondent ID 2357)
>
> I use personal networks and public access datasites to discover data, then I usually have to submit a proposal and get it accepted in order to get access to the data. I have not had the experience of just downloading data directly without going through a permission process. (Respondent ID 1416)

Attending conferences and having discussions within personal networks are the most frequent ways of mobilizing social connections to discover data (Figure 12). While personal networks remain important in actually gaining access to data, contacting data authors directly is the most often reported method for accessing data. Forming new collaborations with data creators also appears to be more important in accessing data than in first discovering them. These patterns hold, regardless of the types of data that respondents need or their intended use for the data, although there are some disciplinary differences in the percentage of respondents discovering or accessing data via conference attendance or forming new collaborations. The need to use personal connections in accessing data also reflects the finding that access remains the largest hurdle for participants.

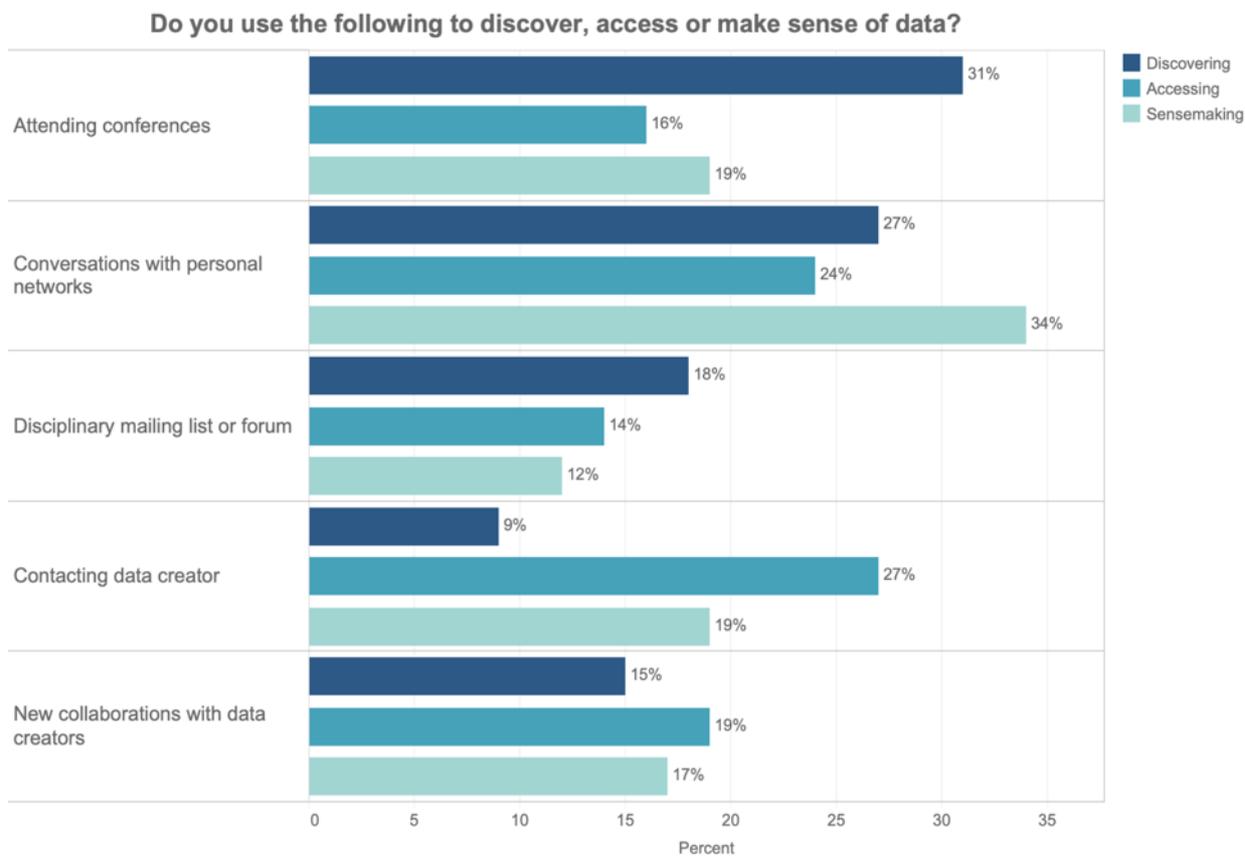

Figure 12. How respondents make use of social connections in: discovering data (n=3311), accessing data (n=3589) and making sense of data (n=3031) . Percent denotes percentage of responses for each option; multiple responses possible.

Social connections are a two-way street. Not only do respondents seek data from their networks, but experts also receive data without solicitation or in exchange for their knowledge. Participants receive data from individuals both within and outside of their domains of expertise.

> Specialist organizations often seek my help and their members send me data. I write to schools to invite them to participate in the experimental work I do and I analyze their data for them and send them reports and suggestions to overcome the difficulties that I observe. (Respondent ID 1242)

> I am an expert in statistics. I didn't have to find the data; the researchers that owned the data find me. I have published a few original articles not being part of my field of expertise as a co-author - as the researcher responsible for the statistical analysis. (Respondent ID 3253)

*Via "mediated" search*

For some respondents, actually locating and accessing data is a mediated process, mediated not through the work of information professionals (although this sometimes happens - see Figure 10), but rather through the literature and through personal connections. Numerous respondents first discover or encounter data via an "intermediary" source - an article, a conversation with a colleague; they then turn to another source - a data repository, Google - to search specifically for the known data.

> I generally do not search for data blindly but I would normally know that it already exists through some previous interaction (reading scientific publication, personal communication). (Respondent ID 679)

General search engines (e.g. Google) can also serve in intermediary roles, as respondents use them not only in order to find data themselves, but also to locate data repositories. These two practices - using Google for known-item searches and to locate repositories rather than data - could contribute to the fact that 38% of respondents who use general search engines found their searches to be either successful or very successful. However, the majority of respondents using general search engines reported mixed success (55%), with 7% being rarely or never successful, perhaps reflecting the higher failure rate in general in academic search compared to general web searching (Li, Schijvenaars, & de Rijke, 2017).

*Via specific searches plus casting a wide net*

Much of data searching is very specific. Participants rely on particular, known data repositories and sources. Respondents have specific requirements and search parameters, and seek data for specific purposes and goals, as is evidenced in our typology of data (re)use (Table 4). This is in contrast to literature searching, where participants report using cross-disciplinary sources, such as the Web of Science or Scopus, and where the goal is often to cast a wide net to discover ideas for use in theory or concept development.

> When I search for data I am pretty focused on finding only data sets that I need for a specific purpose. When I search for literature I read papers that are only peripherally related to the subject but they help me formulate new ideas. (Respondent ID 2128)

> I tend to search for data by specifying parameters e.g. geographical and date coverage, or by looking for data created by a specific organization. My literature searches are more general and don't have so many search filters applied. (Respondent ID 3688)

In contrast, searching for data is haphazard and less systematic than literature search for many respondents, requiring researchers to cast a wide net to discover distributed data.

> It [data search] is a little more haphazard, as I am not as comfortable with finding data. Some of this stems from my not knowing "the" sources, but some of it is also because the finding tools are not yet available. Many times, it is a "try and see" approach. (Respondent ID 3803)

> With scientific literature I know for sure where to look for it, before I start the search. In other words, sources are known to me and do not change for years. With data it is always not so. I may find data in unexpected places. (Respondent ID 3152)

***By building new practices***

This state of development causes search practices to be in flux, as individuals figure out how best to find and access the data that they need. In contrast to their practices for searching literature, their data search practices are still in formation.

> Finding academic literature is part of everyday practice. The processes for finding literature are well-established and institutionally-supported. If I need to find data I have to establish my own process to locate where they are held, get permission from owners, agree access rights etc. (Respondent ID 696)

4.4 RQ 4: How are people evaluating and making sense of data for (re)use?

***By using varied evaluation criteria and sensemaking strategies***

Respondents require a variety of information about the data and make use of different sensemaking strategies (Figure 13). Eighty-nine percent of respondents reported that information about data collection conditions and methodology was important or extremely important in their decisions; information about data processing/handling as well as topical relevance were also ranked highly (Figure 13a). The ease (or difficulty) of accessing data is also very important to 73% of participants. While respondents take the reputation of the data creator into consideration, with 62% of respondents indicating this is important or extremely important, the reputation of the source of the data (e.g. the repository or journal) appears to be slightly more important, as 71% of respondents identified the reputation of the source as being important or extremely important. This is further evidenced in Figure 16a; 61% of respondents selected the data creator's reputation as being important/extremely important in establishing trust in secondary data, as compared to 81% who identified the reputation of the source as being key to developing trust.

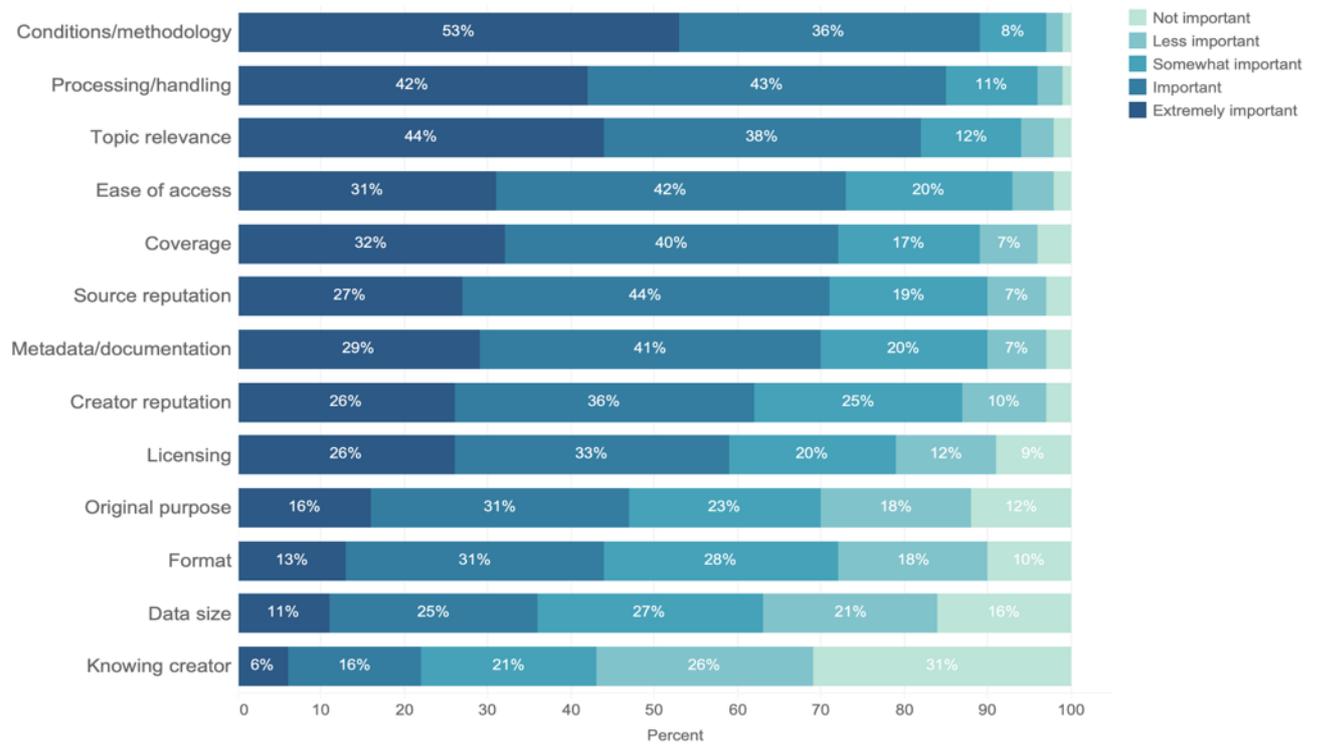

Figure 13a) Information used in evaluating data for reuse (n=1677). Percent denotes percentage of respondents.

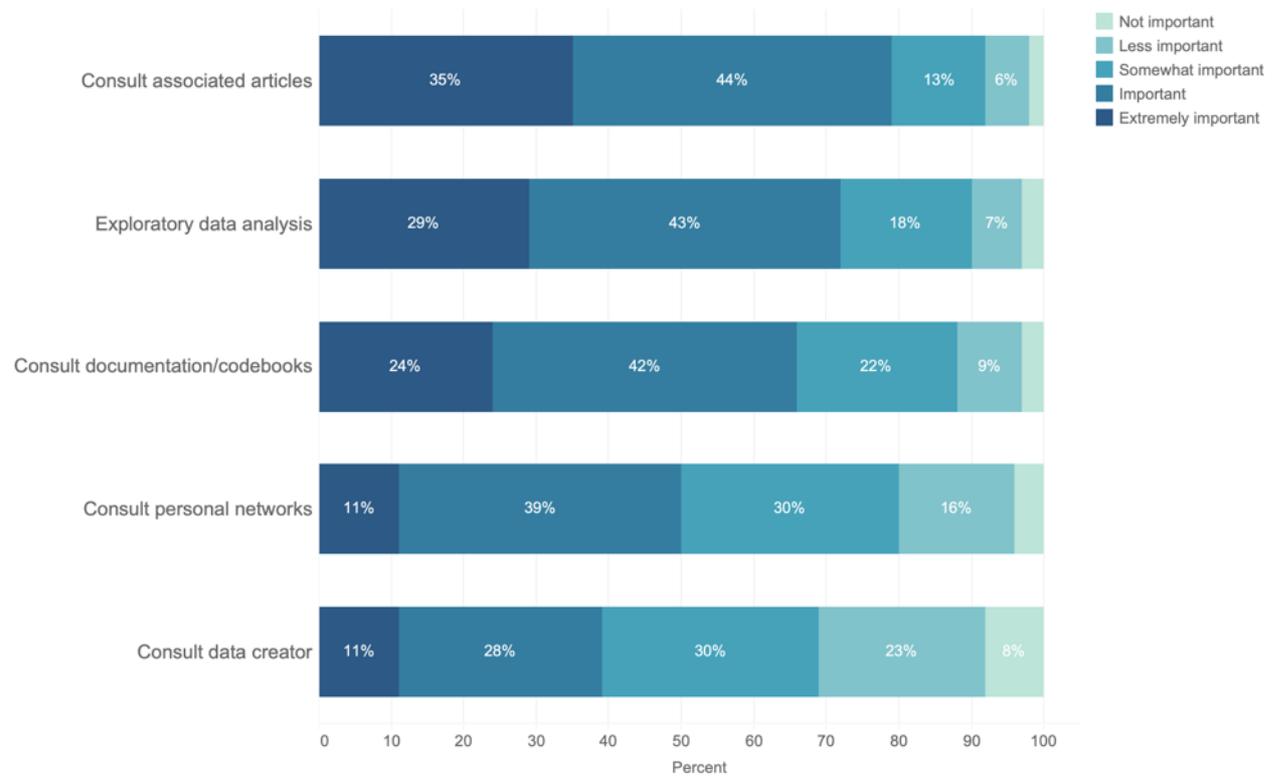

Figure 13b) Sensemaking strategies (n=1677). Percent denotes percentage of respondents.

Other information identified in open responses includes the timeliness of data, prior usage, and the cost of obtaining data, which can determine the type of research that is pursued.

> Patent data is free to access. Data on company deals and revenue can sometimes be paid. That requires seeking research funding and typically delays the process. Using publicly free data is quicker. (Respondent ID 3220)

The academic literature plays a key role not only in discovering data, but also in understanding them. Respondents consult associated articles, as well as data documentation and codebooks (Figure 11b). Nearly three-quarters of respondents report engaging in exploratory data analysis, i.e. statistical checks or graphical analysis. Participants also report triangulating data from multiple sources as a way of understanding and determining the validity of data (e.g. Respondent IDs 3131, 2444, 1949).

*By using social connections in sensemaking*

Fifty percent of participants reported conversations with personal networks as being key to making sense of data. Conversations with networks are used more often in sensemaking than in either discovering or accessing data (Figure 12). Contacting data creators to make sense of data does not happen as frequently as discussions with personal networks.

Respondents also attend conferences and form new collaborations to make sense of data (see Figure 12). Some variations in the pattern in Figure 12 for sensemaking exist across disciplinary domains (e.g. see Figure 14), although engaging in conversations with personal networks is almost always chosen most frequently. These variations are likely the result of different disciplinary norms and infrastructures, which influence patterns of collaboration and communication (e.g. the role of conference attendance and publishing norms, or the existence of disciplinary mailing lists and forums, i.e. in computer science).

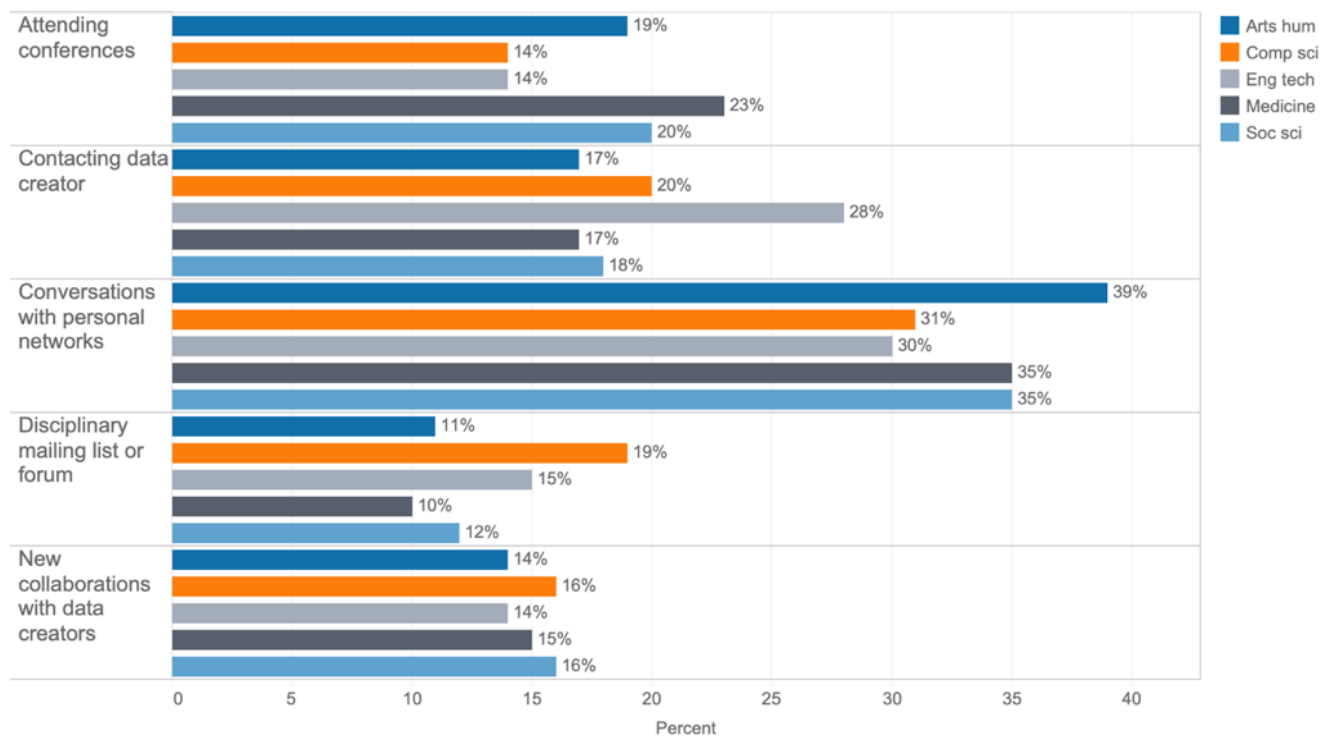

Figure 14. Social strategies of sensemaking in five disciplines from the disciplinary subset: arts and humanities (n=64), computer science (n=94), engineering and technology (n=109), medicine (n=158) and social science (n=155). Percent denotes percentage of responses; multiple responses were allowed.

***By using different contextual information for different purposes***

Different data uses are associated with needing different information about data. Figure 15 presents significant associations, detected using the statistical test for multiple marginal independence described in the Methodology section, between data uses and evaluation criteria. An association detected with this method indicates that responses to the survey question about evaluation criteria are correlated with responses to the question about data uses. In Figure 15, we classify the evaluation criteria presented in Figure 13a into content-related information (e.g. data collection methods and conditions, the relevance of data to a topic, the exact coverage of the data), structure-related information (format, size, the existence of detailed documentation and metadata), access-related information (ease of access, licensing) and social information (reputation of data creator and source, knowing the data creator). We then plot significant associations which exist between uses in the (re)use typology and these evaluation criteria.

This analysis allows us to begin to identify the types of information needed by respondents in different research phases. It also allows us to identify gaps. Most of the detected associations occur between content-related information and data uses in the project creation/preparation or analysis/sensemaking stages of our (re)use typology. The fewest number of significant associations were detected with calibration and benchmarking. Only one association was found for structure-related information, between teaching and data format; source reputation was also correlated with format and idea generation. The importance of information about how data are processed and handled span uses in all of the research phases; having access to detailed and complete metadata or documentation was found to be associated with experimenting with new techniques and methods, with integration, and with making comparisons.

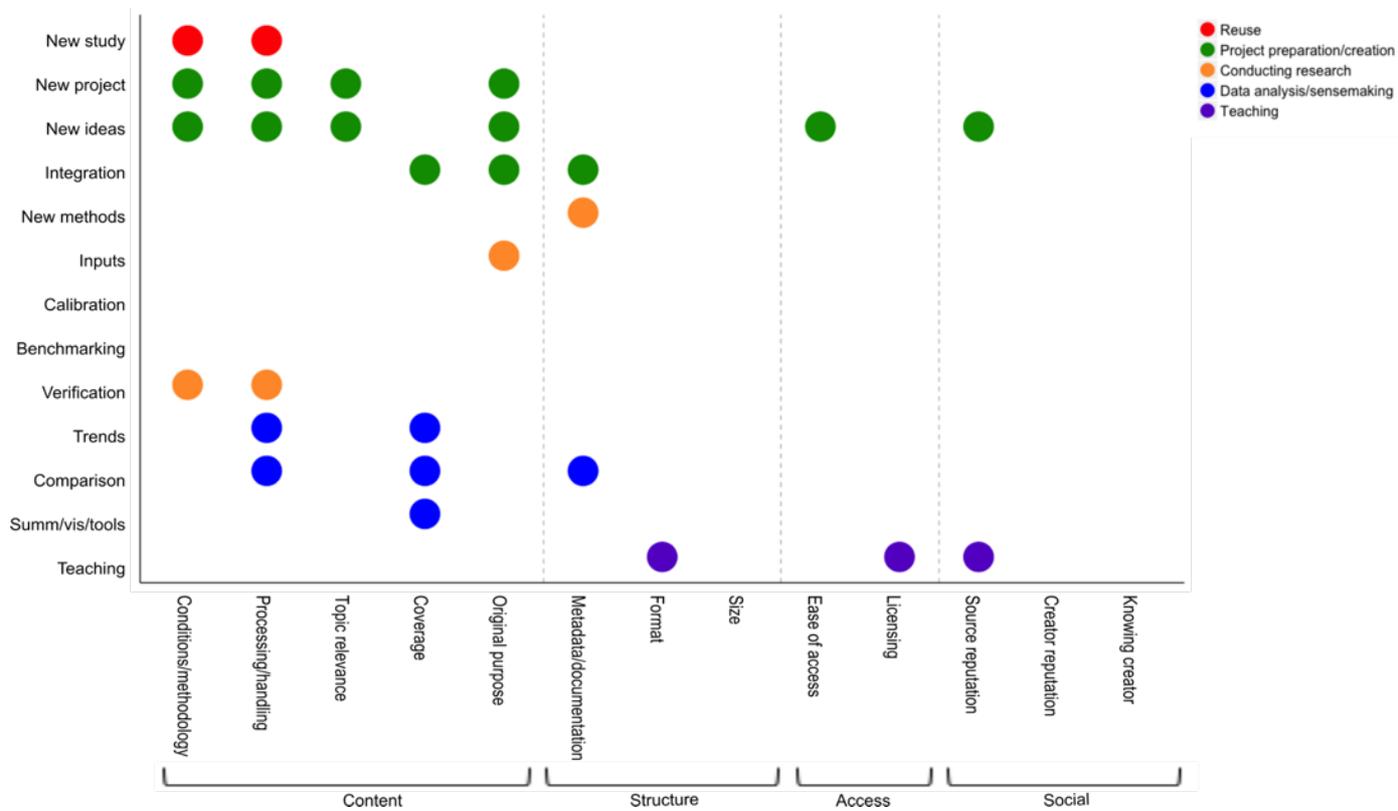

Figure 15. Significant associations between types of data use and information important in evaluating data. Associations detected using adjusted Bonferroni test for multiple marginal independence (significance level: $p < 0.05$). Colors represent phases of (re)use typology. Classifications of evaluation criteria marked with brackets.

### *By establishing trust and data quality*

The transparency of data collection methods, followed by the reputation of the source and a minimum of errors are critical in trust development (Figure 16a).

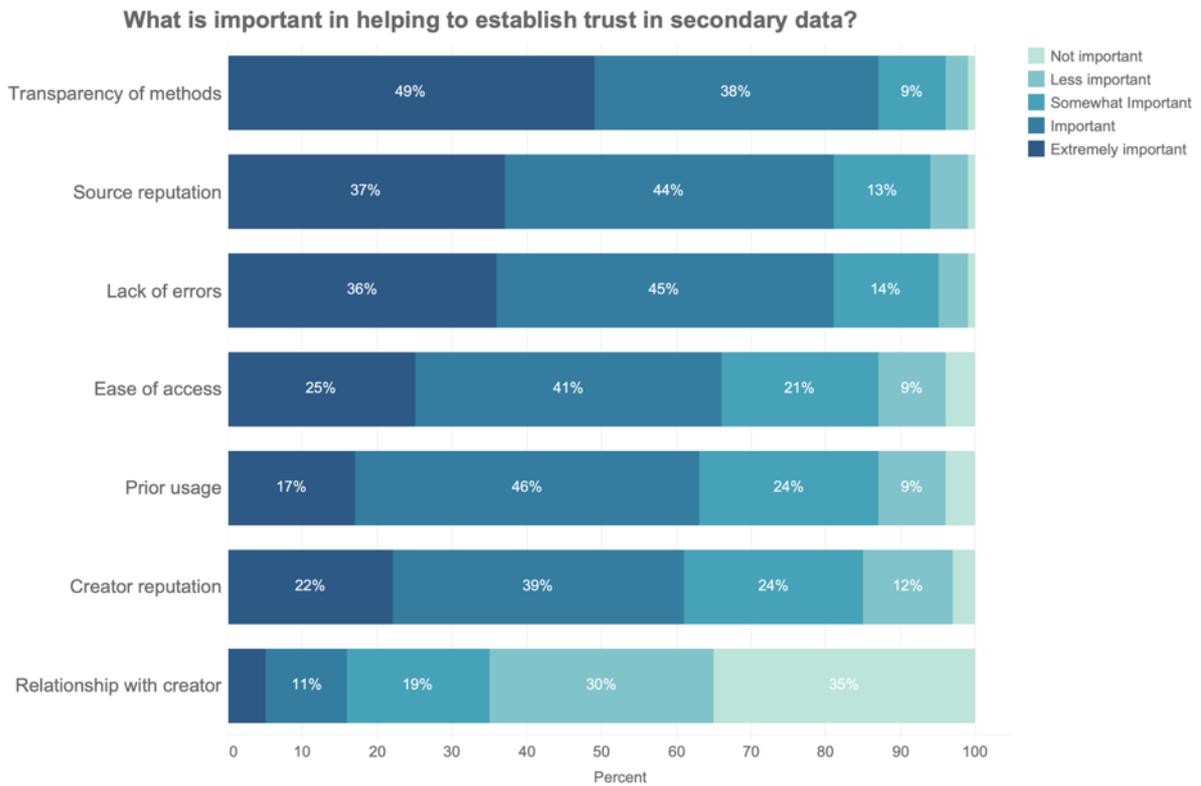

Figure 16a). Importance of criteria used to establish **trust** in secondary data (n=1677). Percent denotes percentage of respondents.

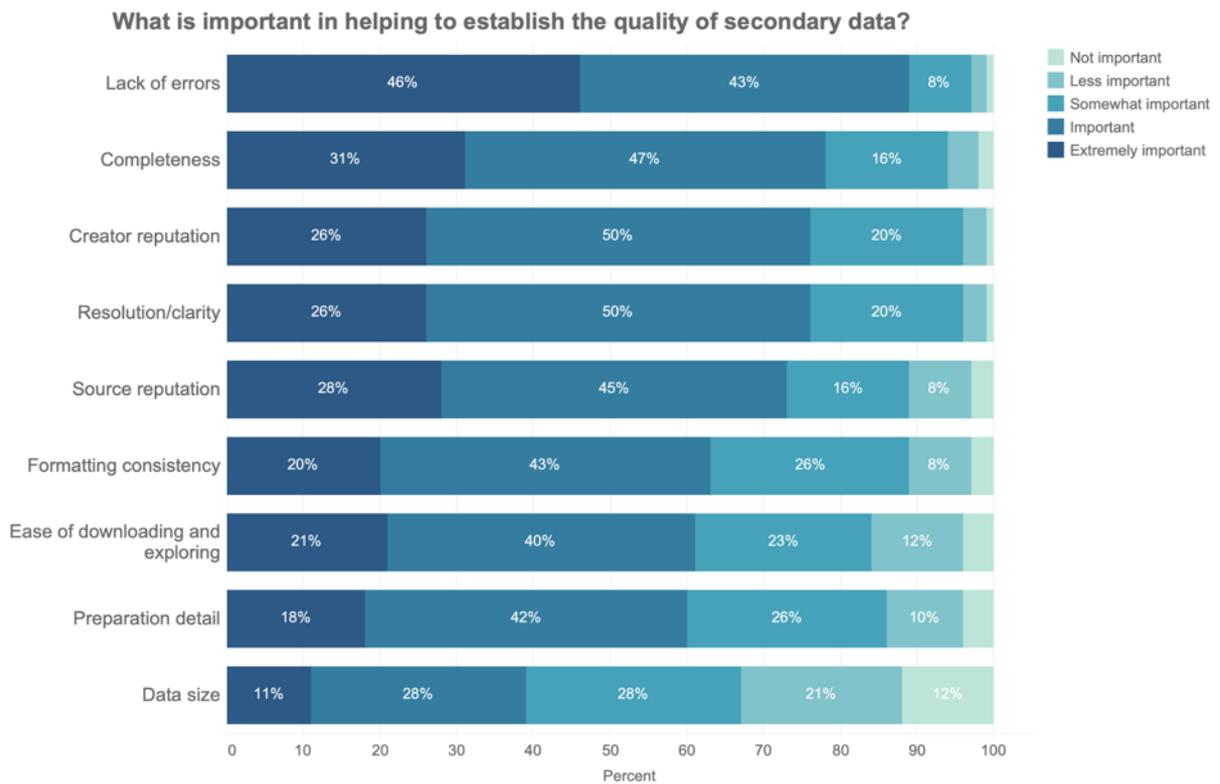

Figure 16b) Importance of criteria used to establish **quality** of secondary data (n=1677). Percent denotes percentage of respondents.

In some disciplines, a completely error-free dataset may actually raise suspicions, as it may indicate that the data have been tampered with or manipulated.

> Lack of errors would not necessarily help establish trust - errors are normal, so a perfect dataset without errors might be a fabricated dataset. (Respondent ID 3970)

> If you know the field you also know what to look for with respect to unreliable data. Sometimes the occasional error actually speaks to the reliability of a dataset: It indicates a person was involved somewhere in data entry. (Respondent ID 1648)

Establishing data quality also depends heavily on the absence of errors and data completeness. Both developing trust and determining quality involve social considerations (Yoon, 2017, Faniel & Yakel, 2017). Although respondents across disciplines consistently ranked having a personal relationship with the data creator as being unimportant in establishing trust, they still weigh other "social" factors - i.e. thinking about human involvement in data creation or the reputation of the source - in their decisions. The reputation of the data creator appears to be more important to respondents when evaluating data quality than in trust development.

## 5. Discussion

We identify and apply four analytical themes to further discuss our findings about practices of data seeking and reuse. We also consider each theme's relation to recent efforts to increase the discoverability of research data before concluding by suggesting future areas for both practical and conceptual work.

### 5.1 Communities of use

The term "community" is often used without considering how communities are formed or their exact composition. Community boundaries are shifting and porous, rather than fixed and stable, and individuals often belong to multiple communities simultaneously (Birnholtz & Bietz, 2003).

We see this clearly in our data. Although communities are typically thought of in terms of disciplinary domains, more than half of our respondents identified with multiple disciplines. Fifty percent also indicated needing data outside of their domains of expertise, perhaps reflecting funders' efforts to promote interdisciplinary research (Allmendinger, 2015). Data communities can also be thought of in terms of the type of data that a particular group uses (Cooper & Springer, 2019; Gregory et al, 2019b). However, we show here that respondents need multiple types of data for their work, and that these data needs can be difficult to classify in broad terms.

Communities can form around particular methodologies and ways of using and working with data (Leonelli & Ankeny, 2015), as is the case, e.g. in the digital humanities and sociology or economics (Levallois, Steinmetz, & Wouters, 2013). Our (re)use typology (Table 4) allows for conceptualizing data communities in terms of broad uses of data, e.g. using data in conducting research, and in more specific terms, e.g. using data for calibration. We also found initial signals that particular data uses are associated with needing certain information about data (Figure 15). Content-related metadata, such as information about collection conditions and methodologies, is important for our respondents in preparing for new projects; structure-related metadata, e.g. format, is important in teaching. We saw a similar relationship, particularly for teaching data science, in our qualitative data. This suggests another way of conceptualizing data seeking communities - by broad research approaches. Individuals relying on data science techniques, no matter their discipline, may rely more on structure-related information when evaluating data for reuse; content-specific considerations could be more important for more traditional research approaches.

Open data policies and guidelines recognize the importance of communities, but often equate communities with disciplinary domains. The FAIR data principles, e.g. (Wilkinson et al., 2016), call for the use of domain-relevant community standards as well as relevant attributes to facilitate findability and reuse. Our analysis encourages a multi-dimensional way of thinking about communities, recognizing that community-relevant metadata can be defined by considering other factors (e.g. data use) in conjunction with disciplinary domains.

## 5.2 Interwoven practices

Data discovery is interwoven with other (re)search practices, particularly searching for academic literature. Roughly eighty percent of respondents stated that their practices for finding data and literature are either sometimes or always the same. The academic literature itself is the go-to source for finding data for the majority of participants. Despite the immature state of data citation practices in many disciplines (Robinson-Garcia et al, 2016), respondents use a strategy common in literature searching - following citations - to locate data from the literature.

Data citation is not equivalent to bibliographic citation (Borgman, 2016). Most data citations indicate some type of data "usage" (Park & Wolfram, 2017), but little is known about why people cite data or the details of how data have been used in a work (Silvello, 2018). This presents an additional challenge for people seeking data to use for a particular purpose. A citation model that typifies how data have been used could potentially facilitate data discovery and evaluation practices that begin by following data citations, as well as add value to the multiplicity of data uses that we observed.

## 5.3 Social connections

Discovering and accessing data are also mediated by personal networks. Respondents find out about data from their connections and then hunt the data down digitally. This process also occurs in reverse - respondents find data digitally and then access the data by personally contacting data creators. The use of social connections in discovery and sensemaking is intertwined with discipline-specific practices of communication (e.g. the role of conference attendance) and collaborations.

Participants identify using social connections as an important difference between searching and accessing data, as opposed to literature. This difference could be a result of the fact that infrastructures to support data search and access are still in development. It could also be due to the complexity of the sociotechnical issues surrounding data access. Researchers question which data to make available for whom (Levin and Leonelli, 2017), and sensitive data containing private information about participants cannot be made openly accessible.

Accessible data, as defined by the FAIR principles, do not necessarily need to be open data (Mons, et al., 2017). Access to data can be mediated by automated authorization protocols (Wilkinson et al, 2016), but automatic denials of access may not mean that data are completely closed to a human data seeker. Researchers can still contact data authors directly if access is denied to learn more about restrictions and possibly form collaborations that would enable reuse (Gregory et al, 2018). Our results also show that ease of access is a top consideration in using data, especially during early phases of research (Figure 15). As certain data become easier to seamlessly and automatically access, other data, those that are more challenging to access, will likely not be used as often, which will shape the research that is or is not pursued.

## 5.4 Practices in flux

Practices and infrastructures are closely linked (Shove, Watson, & Spurling, 2015); this is especially true for practices of data discovery and reuse. We see this in the tension that we found between specific and haphazard search practices. For some respondents, data infrastructures are still in a state of development, which requires casting a wider net to locate appropriate sources. For others, finding data involves going directly to a particular, well-known data repository in the field.

Data infrastructures consist of assemblages of policies, people, technology and data (Borgman et al., 2015, Edwards, 2010). As data are described in more standardized ways, repositories will have different methods of structuring data, be linked to other data and repositories more seamlessly, and will build new services. These services and linkages will change how data seekers interact with and discover data.

Innovations combining new technologies with existing practices will not only alter current practices, but will also bring new considerations to the forefront. Executable papers (e.g. Gil et al, 2016) where readers can interact with data directly, build on linkages between literature and data searching as well as the importance of exploratory data analysis in sensemaking. They also blur the line between where a paper ends and data begins. As the boundaries between data and papers become less defined, the importance of archiving those data in sustainable ways (e.g. Vander Sande, Verbogh, Hochstenbach, & Van de Sompel, 2018) and questions of properly citing data creators, rather than paper authors, will become more visible.

# 6. Conclusion

In summary, we have examined the data needs, uses, discovery strategies and sensemaking practices for the 1677 respondents to our survey by presenting an initial quantitative analysis and by drawing on the qualitative survey data. Possible practical applications for this work are many. We conclude by highlighting some key takeaways from our analysis and draw attention to their potential applications, in particular for designers of data discovery systems and managers of data repositories.

6.1 Diversity is normal, not abnormal

Past in-depth ethnographic work has documented the diversity of data practices in particular research groups and projects within disciplines such as astronomy, environmental engineering and biomedicine (see Borgman, Wofford, Golshan, Darch & Scroggins, 2019). Both our quantitative and qualitative results show that this multiplicity is not limited to these specific communities. Rather than being the exception, a diversity of data needs, data uses and data sources appears to be the rule. We also see signals of other forms of diversity in our results, finding, e.g. that data needs and search practices are both specific and broad, that data uses are spread across phases of research and that practices of data discovery are intermingled with other search practices.

We have suggested elsewhere that data discovery systems implement a variety of differentiated search interfaces, including visual and graphical navigation systems, for different users (Gregory et al, 2019a), which would allow data seekers freedom to explore data in different ways. The diversity we observe here also supports the creation of flexible, interlinked designs for data discovery systems and repositories with different levels of specificity. Linking diverse data across locations via standardized approaches is gaining significant momentum (see, e.g. Wilkinson et al, 2016), and would pave the way for general data search engines and federated search efforts. Domain agnostic data search engines (i.e. Google Dataset[7] or DataSearch[8]) allow data seekers to search for data broadly before being directed to data repositories, where more specific searching and exploration is often possible. This puts an onus on repositories, however, as they become responsible for implementing a new generation of tools supporting more specific search and sensemaking activities within their environments (i.e. generous interfaces (Whitelaw 2015; Mutschke et al. 2014)).

Data search engines could also feasibly help searchers looking for data outside of their domain to better identify potential repositories of interest. Our findings also suggest the need to integrate data search tools with literature databases and data management tools, which we also suggested based on our earlier interviews with data seekers (Gregory et al, 2019a). Figure 9 shows that this integration may be more relevant for certain disciplines than for others.

6.2 Communities of use as an entry point to design

---

[7] https://toolbox.google.com/datasetsearch
[8] https://datasearch.elsevier.com/#/

While we suggested differentiated interfaces for different users in the past, the multiplicity of practices we observed makes it challenging to design tools based on "user profiles;" there is also often a gap between the user profiles which designers imagine and the actual users themselves (see Wyatt, 2008). Both our quantitative and qualitative data suggest that searching for data is purpose-driven; researchers look for data for a specific purpose or use. This finding is supported by Koesten et al, who use interviews with data professionals and log analyses to suggest that data search is task-dependent (2017). We suggest here also that data communities can be conceptualized by uses of data and show in Figure 7 that these uses, although spread across research phases and not limited to a particular point in time (see also Pasquetto et al, 2019), appear to be limited to a core set. This set of data uses could provide an entry point for design, allowing data seekers to search for or filter by data used for these specific purposes.

The set of uses we identify is general and needs further research to test its comprehensiveness. Performing cluster analyses using the data from our survey may be one way to further identify and validate the *communities of use* which we propose. It is also likely that repositories may be able to identify their own specific communities of use. Identifying these will require innovative approaches by repositories and communication with both data depositors and consumers (see section 6.4).

### 6.3 Metadata to support sensemaking and reuse

Metadata plays a key role in facilitating data reuse (Pasquetto et al, 2019; Mayernik, 2016; Mayernik & Acker, 2017); however, the general metadata needed for discovery, is often not rich enough to support the sensemaking needed for reuse (Zimmerman, 2007). In Figure 13 and 16, we identify common evaluation criteria and considerations in developing trust and quality which could inform the development of metadata, at both broader and more specific levels, to support making decisions about using data. Figure 15, in particular suggests that certain evaluation criteria can be used to support certain types of data uses, although additional testing needs to be done to validate the associations we detected and to investigate how the factors identified in Figure 13 and 16 vary by use or disciplinary domain.

### 6.4 Tenacity and value of the social

An increasing amount of quantitative (e.g. Digital Science, 2019) and qualitative work (e.g. Yoon, 2017) shows that using secondary data requires communication and collaboration. Our findings demonstrate the importance of social interactions in discovery, access, and sensemaking, in particular showing that researchers rely on conversations with personal networks to make sense of secondary data. We believe that these social interactions will continue to be important in data discovery and reuse, and should be seen as something to support when designing systems and repositories.

Our earlier suggestions, i.e. ranking datasets by social signals or integrating offline and online interactions around data, still hold promise (see Gregory et al, 2019a). We could also see a role for an expanded metadata schema as a way to open a conversation between data producers and multiple data consumers within the context of the data themselves at the repository level. In such a system, both data producers and data consumers could contribute information about the data to specific fields. Data producers could describe their own use of the data as well as their concerns about what potential reusers need to know about the data before reusing them. Data users could describe how they used or plan to use the data, as well as pose questions about the data to other users or to the producer. Such an implementation would result in a layered metadata record, including general metadata needed for discovery, metadata with common elements supporting sensemaking (e.g. those identified in Figure 13), and a layer of interactive, or communication-based metadata. In the future, we envision a co-evolution of metadata schemes driven both by designated communities and repository managers. Such a dynamic way to co-construct metadata schemas and indexing could enable repositories to identify new communities of use.

Our quantitative data provides the opportunity to more deeply probe and test the results we present here. Creating multi-level models to further explore the influence of data uses, data types, or disciplinary domains could be a fruitful next step, as could investigating methods to test the generalizability of our data to broader populations. This future work, in conjunction with the results in this paper, could also inform deeper theoretical work. Designing useful, sustainable tools and services requires considering the interconnections between different practices, infrastructures and communities that we have begun to investigate. Further conceptual work needs to be done to highlight these connections in a way that can be easily communicated and that can practically inform design.

## Acknowledgments

We are very grateful to Ricardo Moreira for his advice and help in organizing, scripting, distributing and managing the survey and to Helena Cousijn for her advice in designing the survey. We would also like to thank Natalie Koziol for her assistance with using the MRCV package. This work is part of the project *Re-SEARCH: Contextual Search for Research Data* and was funded by the NWO Grant 652.001.002

## Author Contribution Statement
KG created the survey, analyzed the data and wrote the paper. PG provided input and feedback on the survey and data analysis as well as drafts of the paper. AS and SW provided input and feedback on the survey design and earlier drafts of the paper.

**Supplementary material**

**Appendix A: Survey questionnaire**

**Introduction**

This study investigates how participants locate and evaluate data they do not create themselves.

The survey consists of three main sections:

• Part 1: Data Needs

• Part 2: Finding Data

• Part 3: Evaluating Data

Our funding comes from the Netherlands Organization for Scientific Research (NWO). The study is part of a collaborative research project between researchers at the Data Archiving and Networked Services (DANS), the University of Amsterdam, the Vrije Universiteit Amsterdam and Elsevier.

By clicking on the below button to start the survey, you indicate your consent to participate in this research. You can read more about the survey and what will be done with the data here (this will launch a new window).

Thank you for your participation.

Please click >> button to indicate consent to participate and to begin the survey.

**Survey Questions**

**Part 1: Data Needs**

**Q1: Which of the following best describes you?**

*Please select one answer*

○ Researcher
○ Student

❍ Librarian, archivist or research/data support provider
❍ Manager
❍ Other. Please specify ____________

**Q2: Please describe the secondary data that you (might) need. (We define secondary data as data that you do not create yourself).**

*Please write your answer in the box below:*

[                                                                                              ]

**Q3: Please select the options that describe the secondary data that you (might) need.**

*Please select all that apply*

❑ Observational or empirical (e.g. sensor data, survey data, interview transcripts, sample data, neuroimages, ethnographic data, diaries)
❑ Experimental (e.g. gene sequences, chromatograms, toroid magnetic field data)
❑ Simulation (e.g. climate models, economic models)
❑ Derived or compiled (e.g. text and data mining, compiled database, 3D models)
❑ Other, Please specify ____________

**Q4: Why do you use or need secondary data?**

*Please select all that apply*

❑ As the basis for a new study
❑ To calibrate instruments or models
❑ For benchmarking
❑ To verify my own data
❑ As model, algorithm or system inputs
❑ To generate new ideas
❑ For teaching/training
❑ To prepare for a new project or proposal
❑ To experiment with new methods and techniques (e.g. to develop data science skills)
❑ To identify trends or make predictions
❑ To compare multiple datasets to find commonalities or differences
❑ To create summaries, visualizations, or analysis tools
❑ To integrate with other data to create a new dataset
❑ Other. Please specify ____________

**Q5: Have you ever used data outside of your area of expertise?**

*Please select one answer*

❍ Yes
❍ No

**Q5a: How did you find this data?**

*Please write your answer in the box below:*

[                                                                                              ]

**Part 2: Finding Data**

**Q6: When you need data, who finds it for you?**

*Please select all that apply*

❏ I find it myself
❏ Graduate student
❏ Research support professional (e.g. librarian, archivist, data or literature manager)
❏ Someone else in my personal network (e.g. peers, collaborators, mentors)
❏ Other. Please specify ____________

**Q7: How frequently do you use the following to find data?**

*Please select one answer per row*

|  | Often | Occasionally | Never |
|---|---|---|---|
| Multidisciplinary data repositories |  |  |  |
| Discipline-specific data repositories |  |  |  |
| Governmental agencies and websites |  |  |  |
| Personal networks (e.g. colleagues, peers) |  |  |  |
| Academic literature (e.g. journal articles, conference proceedings |  |  |  |
| Code repository (e.g. GitHub) |  |  |  |
| General search engines (e.g. Google) |  |  |  |
| Professional associations |  |  |  |
| Data specific search engines |  |  |  |
| Commercial sources |  |  |  |
| Consultation with research support professionals (e.g. librarians, archivists or data managers) |  |  |  |

**Q7_open: Please specify any other resources that you use to find data:**

*Please write your answer in the box below:*

**Q7a: Which statement(s) describe how you discover data using the academic literature?**

*Please select all that apply*

❏ I search the academic literature with the goal of finding data.
❏ I find data serendipitously while reading articles or performing literature searches.
❏ I follow citations and references in the literature to datasets.
❏ I extract and use data from the literature directly (e.g. from tables, graphs, or instrument specifications and parameters)
❏ Other. Please specify ____________

**Q7b How successful are you at finding data with a general search engine (e.g. Google)?**

*Please select one answer*

○ Very successful
○ Successful
○ Sometime successful, sometimes not
○ Rarely successful
○ Not successful

**Q8: How frequently do you find data in the following ways?**

*Please select one answer per row*

| | Often | Occasionally | Never |
|---|---|---|---|
| By actively searching for data in an online resource | | | |
| Serendipitously, when searching for something else (e.g. when looking for journal articles or news) | | | |
| Serendipitously, when NOT actively looking for something else (e.g. via an email notice or interaction with a colleague) | | | |
| In the course of sharing or managing my own data | | | |

**Q9: Please indicate if you use the following to discover, access, or make sense of data.**

*Please select all that apply*

| | Q10a - Discover | Q10b - Access | Q10c - Making sense of data |
|---|---|---|---|
| Conversations with personal networks (e.g. colleagues, peers) | | | |
| Contacting the data creator | | | |
| Developing new academic collaborations with data creators | | | |
| Attending conferences | | | |
| Disciplinary mailing lists or discussion forums | | | |

**Q10: Do you discover data differently than how you discover academic literature?**

*Please select one answer*

❍ Yes
❍ Sometimes
❍ No

**Q10a: How is your process for finding data different than your process for finding academic literature?**

*Please write your answer in the box below:*

**Q11: How easy is it to find data?**

*Please select one answer*

❍ Easy
❍ Sometimes challenging
❍ Difficult

**Q11a: Why is it challenging to find the data that you need?**

*Please select all that apply*

❑ The data are not accessible (e.g. behind paywalls, held by industry).
❑ I don't know where or how to best look for the data.
❑ The data are located in many different places.
❑ The data are not digital.
❑ Online search tools are inadequate.
❑ I do not have the personal network needed to find or access the data.
❑ Other. Please specify ____________

## Part 3: Evaluating Data

**Q12: Please indicate the importance of the following information when deciding whether or not to use secondary data.**

*Please select one answer per row*

|  | Extremely important | Important | Somewhat important | Less important | Not important |
|---|---|---|---|---|---|
| Data collection conditions and methodology |  |  |  |  |  |
| How data has been processed and handled |  |  |  |  |  |
| Reputation of data creator |  |  |  |  |  |
| Personally knowing the data creator |  |  |  |  |  |
| Reputation of data source (e.g. repository or journal) |  |  |  |  |  |
| Detailed and complete metadata and documentation |  |  |  |  |  |
| Data size |  |  |  |  |  |
| Data format |  |  |  |  |  |
| Licensing/copyright conditions |  |  |  |  |  |
| Correct coverage (time, location, population, etc.) |  |  |  |  |  |
| Original purpose of the data |  |  |  |  |  |
| Ease of access |  |  |  |  |  |
| Topic relevance |  |  |  |  |  |

**Q12_open: Please specify any other information you consider when deciding whether to use or not secondary data.**

*Please write your answer in the box below:*

**Q13: How important are the following strategies in evaluating and making sense of data?**

*Please select one answer per row*

|  | Extremely important | Important | Somewhat important | Less important | Not important |
|---|---|---|---|---|---|
| Consulting associated journal articles |  |  |  |  |  |
| Consulting data documentation and codebooks |  |  |  |  |  |
| Consulting the data creator |  |  |  |  |  |
| Consulting personal networks (e.g. colleagues, peers) |  |  |  |  |  |
| Exploratory data analysis (e.g. statistical checks, graphical analysis) |  |  |  |  |  |

**Q13_open: Please specify any other strategies you consider to evaluate and make sense of data.**

*Please write your answer in the box below:*

**Q14: Please indicate the importance of the following in helping you to establish trust in secondary data.**

*Please select one answer per row*

|  | Extremely important | Important | Somewhat important | Less important | Not important |
| --- | --- | --- | --- | --- | --- |
| Others' prior usage of the data |  |  |  |  |  |
| Reputation of source (e.g. repository, journal) |  |  |  |  |  |
| Reputation of data creator |  |  |  |  |  |
| Transparency in data collection methods |  |  |  |  |  |
| Lack of errors |  |  |  |  |  |
| Ease of access |  |  |  |  |  |
| Personal relationship with the data creator |  |  |  |  |  |

**Q14_open: Please specify any other important aspects you consider to help establish trust in secondary data.**

*Please write your answer in the box below:*

**Q15: Please indicate the importance of the following in helping you to establish the quality of secondary data.**

*Please select one answer per row*

|  | Extremely important | Important | Somewhat important | Less important | Not important |
| --- | --- | --- | --- | --- | --- |
| Lack of errors |  |  |  |  |  |
| Ease of downloading and exploring data |  |  |  |  |  |
| Data size |  |  |  |  |  |
| Data completeness |  |  |  |  |  |
| Reputation of source (e.g. repository, journal) |  |  |  |  |  |
| Resolution or clarity |  |  |  |  |  |
| Reputation of data creator |  |  |  |  |  |
| Detail or amount of work done to prepare data |  |  |  |  |  |
| Consistency of formatting |  |  |  |  |  |

**Q15_open: Please specify any other important aspects you consider to help establish the quality of secondary data.**

*Please write your answer in the box below:*

[                                                    ]

**Part 4: Demographics**

**You are nearly at the end of the survey. Below are some questions to help us classify your answers.**

**D1: In which subject discipline do you specialize?**

*Please check all that apply.*

- ❏ Agriculture
- ❏ Arts and Humanities
- ❏ Astronomy
- ❏ Biochemistry, Genetics, and Molecular Biology
- ❏ Biological Sciences
- ❏ Business, Management and Accounting
- ❏ Chemical Engineering
- ❏ Chemistry
- ❏ Computer Sciences / IT
- ❏ Decision Sciences
- ❏ Dentistry
- ❏ Earth and Planetary Sciences
- ❏ Economics, Econometrics and Finance
- ❏ Energy
- ❏ Engineering and Technology
- ❏ Environmental Sciences
- ❏ Health professions
- ❏ Immunology and Microbiology
- ❏ Materials Science
- ❏ Mathematics
- ❏ Medicine
- ❏ Multidisciplinary
- ❏ Neuroscience
- ❏ Nursing
- ❏ Pharmacology, Toxicology and Pharmaceutics
- ❏ Physics
- ❏ Psychology
- ❏ Social Science
- ❏ Veterinary
- ❏ Information science
- ❏ Other. Please specify____________

**D2: How many years of professional experience do you have in your field?**

*Please select one answer*

- ○ 0-5
- ○ 6-15
- ○ 16-30
- ○ 31+

**D3: In which county do you currently work?**

- ○ Afghanistan
- ○ Albania
- ○ Algeria
- ○ American Samoa
- ○ Andorra
- ○ Angola
- ○ Anguilla
- ○ Antarctica

- ○ Antigua and Barbuda
- ○ Argentina
- ○ Armenia
- ○ Aruba
- ○ Australia
- ○ Austria
- ○ Azerbaijan
- ○ Bahamas
- ○ Bahrain
- ○ Bangladesh
- ○ Barbados
- ○ Belarus
- ○ Belgium
- ○ Belize
- ○ Benin
- ○ Bermuda
- ○ Bhutan
- ○ Bolivia
- ○ Bosnia and Herzegovina
- ○ Botswana
- ○ Brazil
- ○ British Indian Ocean Territory
- ○ Brunei
- ○ Brunei Darussalam
- ○ Bulgaria
- ○ Burkina Faso
- ○ Burundi
- ○ Cambodia
- ○ Cameroon
- ○ Canada
- ○ Cape Verde
- ○ Cayman Islands
- ○ Central African Republic
- ○ Chad
- ○ Chile
- ○ China
- ○ Christmas Island
- ○ Cocos (Keeling) Islands
- ○ Colombia
- ○ Comoros
- ○ Congo
- ○ Cook Islands
- ○ Costa Rica
- ○ Cote d'Ivoire
- ○ Croatia
- ○ Cuba
- ○ Cyprus
- ○ Czech Republic
- ○ Denmark
- ○ Djibouti
- ○ Dominica
- ○ Dominican Republic
- ○ East Timor
- ○ Ecuador

- Egypt
- El Salvador
- Equatorial Guinea
- Eritrea
- Estonia
- Ethiopia
- Falkland Islands (Malvinas)
- Fiji
- Finland
- France
- French Guiana
- French Polynesia
- French Southern Territories
- Gambia
- Georgia
- Germany
- Ghana
- Gibraltar
- Greece
- Greenland
- Grenada
- Guadeloupe
- Guam
- Guatemala
- Guinea-Bissau
- Haiti
- Heard Island and McDonald Islands
- Holy See (Vatican City State)
- Honduras
- Hong Kong
- Hungary
- Iceland
- India
- Indonesia
- Iran (Islamic Republic of)
- Iraq
- Ireland
- Israel
- Italy
- Jamaica
- Japan
- Jordan
- Kazakhstan
- Kenya
- Kiribati
- North Korea
- Kuwait
- Kyrgyzstan
- Lao People's Democratic Republic
- Laos
- Latvia
- Lebanon
- Lesotho
- Liberia

- Libyan Arab Jamahiriya
- Lithuania
- Luxembourg
- Macau
- Madagascar
- Malawi
- Malaysia
- Maldives
- Mali
- Malta
- Martinique
- Mauritania
- Mauritius
- Mexico
- Micronesia (Federated States of)
- Monaco
- Mongolia
- Montserrat
- Morocco
- Mozambique
- Myanmar
- Namibia
- Nauru
- Nepal
- Netherlands
- Netherlands Antilles
- New Caledonia
- New Zealand
- Nicaragua
- Niger
- Nigeria
- Niue
- Norfolk Island
- Norway
- Oman
- Pakistan
- Palau
- Panama
- Papua New Guinea
- Paraguay
- Peru
- Philippines
- Pitcairn
- Poland
- Portugal
- Puerto Rico
- Qatar
- Reunion
- Romania
- RUSSIA
- Rwanda
- Saint Helena
- Saint Kitts and Nevis
- Saint Lucia

- Saint Vincent and the Grenadines
- Samoa
- Sao Tome and Principe
- Saudi Arabia
- Senegal
- Serbia and Montenegro
- Seychelles
- Sierra Leone
- Singapore
- Slovakia
- Slovenia
- Solomon Islands
- Somalia
- South Africa
- South Korea
- Spain
- Sri Lanka
- Sudan
- Suriname
- Swaziland
- Sweden
- Switzerland
- Syrian Arab Republic
- Taiwan
- Tajikistan
- TANZANIA
- Thailand
- Togo
- Tonga
- Trinidad and Tobago
- Tunisia
- Turkey
- Turkmenistan
- Turks and Caicos Islands
- Uganda
- Ukraine
- United Arab Emirates
- United Kingdom
- United States Minor Outlying Islands
- Uruguay
- USA
- Uzbekistan
- Vanuatu
- Venezuela
- Viet Nam
- Virgin Islands
- Virgin Islands (US)
- Virgin Islands, British
- Wallis and Futuna
- Yemen
- Zambia
- Zimbabwe
- Palestinian Territory, Occupied
- Moldova, Republic of

- ○ Marshall Islands
- ○ Macedonia, The Former Yugoslav Republic of
- ○ Liechtenstein
- ○ Korea, Republic of
- ○ Guyana
- ○ Guinea
- ○ Gabon
- ○ Faroe Islands
- ○ Zanzibar
- ○ Tokelau

**D4: What type of organization do you work for?**

*Please select one answer*

- ○ University or college
- ○ Research institution
- ○ Government agency
- ○ Corporate
- ○ Independent archive or library
- ○ Other. Please specify ____________

**D5: Please indicate how the following people feel about sharing their research data.**

*Please select one answer per row*

|  | Data sharing is strongly encouraged | Data sharing is somewhat encouraged | Data sharing is neither encouraged nor discouraged | Data sharing is somewhat discouraged | Data sharing is strongly discouraged | Don't know/ Not applicable |
|---|---|---|---|---|---|---|
| You |  |  |  |  |  |  |
| The people you work with directly |  |  |  |  |  |  |
| Your disciplinary community |  |  |  |  |  |  |
| Your institution |  |  |  |  |  |  |

**D6: Please indicate how the following people feel about reusing data produced by other people.**

*Please select one answer per row*

|  | Data reusing is strongly encouraged | Data reusing is somewhat encouraged | Data reusing is neither encouraged nor discouraged | Data reusing is somewhat discouraged | Data reusing is strongly discouraged | Don't know/ Not applicable |
|---|---|---|---|---|---|---|
| You |  |  |  |  |  |  |
| The people you work with directly |  |  |  |  |  |  |

|  | Data reusing is strongly encouraged | Data reusing is somewhat encouraged | Data reusing is neither encouraged nor discouraged | Data reusing is somewhat discouraged | Data reusing is strongly discouraged | Don't know/ Not applicable |
| --- | --- | --- | --- | --- | --- | --- |
| Your disciplinary community |  |  |  |  |  |  |
| Your institution |  |  |  |  |  |  |

**D7: Have you ever shared your own research data?**

*Please select one answer*

❍ Yes
❍ No

**D8: Final comments: Do you have anything else that you would like us to know?**

*Please write your comments in the box below:*

[                                                                                           ]

**Additional questions asked to participants selecting "Librarian, archivist or research/data support provider" as their role.**

**L3: Do you use or need secondary data for your own research or to support others?**

*Please select one answer*

❍ For my own research
❍ To support others
❍ For both my own research and to support others

**L4: Who are the people whom you support?**

*Please select all that apply*

❑ Students
❑ Researchers
❑ Industry employees
❑ Other. Please specify ____________

**L5: How do you support people with their data needs?**

*Please select all that apply*

❑ I teach people about data management planning (e.g. through consultations, workshops, etc.).
❑ I teach people how to discover and evaluate data (e.g. through consultations, workshops, etc.).
❑ I find data for people.
❑ I help people to curate their data.
❑ I find literature for people.
❑ Other. Please specify ____________

# Appendix B: P-value tables

| | Observ/empirical | Experimental | Deriv/compiled | Simulation | Other |
|---|---|---|---|---|---|
| **Agricul** | 0.647 | p < .001* | 0.281 | 0.066 | 0.59 |
| **Arts hum** | 1.000 | p < .001* | p < .001* | p < .001* | p < .001* |
| **Astronom** | 0.103 | 0.92 | 0.008 | p < .001* | 0.92 |
| **Biochem** | p < .001* | p < .001* | 0.92 | 0.003 | 0.322 |
| **Biolog** | p < .001* | p < .001* | 0.069 | 0.091 | 0.806 |
| **Busin** | 0.001 | p < .001* | 0.054 | 0.718 | 0.267 |
| **Chem** | p < .001* | p < .001* | 0.647 | 0.624 | 0.446 |
| **Chem Eng** | p < .001* | p < .001* | 1.000 | 0.002 | 0.862 |
| **Comp Sci** | 0.92 | 0.032 | p < .001* | 0.001 | 0.203 |
| **Decis Sci** | 0.275 | 0.138 | p < .001 | 0.001 | 0.752 |
| **Dentist** | 0.39 | 0.023 | 0.296 | 0.11 | p < .001* |
| **Earth Plan** | 0.003 | 1.000 | 0.02 | p < .001* | 0.92 |
| **Econ** | p < .001 | p < .001* | 1.000 | p < .001* | 0.232 |
| **Energy** | 0.087 | 0.001 | 0.862 | p < .001* | 0.603 |
| **Eng Tech** | 0.009 | 0.022 | 0.187 | p < .001* | 0.488 |
| **Environ** | 0.098 | 0.841 | 0.647 | p < .001* | 0.791 |
| **Health Prof** | p < .001 | 0.002 | 0.458 | 0.001 | 0.841 |
| **Immun** | 0.03 | p < .001* | 0.462 | 0.001 | 0.538 |
| **Info Sci** | 0.002 | 0.001 | p < .001* | 0.031 | 0.004 |
| **Math** | 0.145 | 0.45 | 0.029 | p < .001* | p < .001* |
| **Matl Sci** | p < .001* | p < .001* | 0.103 | 0.008 | 0.59 |
| **Med** | p < .001* | 0.187 | 0.025 | p < .001* | 0.475 |
| **Multi** | 0.005 | 0.92 | 0.013 | 0.002 | 0.823 |
| **Neuro** | 0.137 | 0.123 | 0.764 | 0.09 | 0.639 |
| **Nurs** | 0.187 | 0.037 | 0.252 | 0.144 | 0.729 |
| **Pharma** | 0.03 | p < .001* | 0.639 | 0.079 | 0.538 |
| **Physics** | p < .001* | p < .001* | 0.806 | p < .001* | 0.094 |
| **Psych** | p < .001* | 0.006 | 0.458 | 0.001 | 0.862 |
| **Soc Sci** | p < .001* | p < .001* | 0.332 | 0.005 | 0.791 |
| **Vet** | 0.729 | p < .001* | 0.092 | 0.046 | 0.21 |
| **Other** | 0.071 | 0.037 | 0.377 | 0.074 | 0.035 |

Table 5. P-value table for Figure 6: associations between disciplinary domain and needed data. Significance was determined at the p<.05 level with a Bonferroni correction with m=155 . Significant associations are marked with an asterisk and colored in blue.

| | Observ/empirical | Experimental | Deriv/compiled | Simulation |
|---|---|---|---|---|
| **New study** | p < .001* | 0.044 | 0.007 | 0.566 |
| **New projects** | 0.001 | 0.043 | 0.09 | 0.764 |
| **New ideas** | 0.124 | 0.005 | 0.017 | 0.153 |
| **Integration** | p < .001* | 0.512 | p < .001* | 0.003 |
| **New methods** | 0.33 | p < .001* | p < .001* | p < .001* |
| **Inputs** | 0.135 | p < .001 | p < .001* | p < .001* |
| **Calibration** | 0.242 | p < .001* | p < .001* | p < .001* |
| **Benchmark** | 0.017 | 0.095 | p < .001* | p < .001* |
| **Verification** | 0.251 | p < .001* | p < .001* | p < .001* |
| **Trends** | p < .001* | 0.888 | p < .001* | p < .001* |
| **Comparison** | p < .001* | p < .001* | p < .001* | 0.002 |
| **Summ/vis/tools** | p < .001* | 0.292 | p < .001* | p < .001* |
| **Teaching** | p < .001* | 0.624 | 0.151 | 0.01 |
| **Other** | 0.003 | 0.095 | 0.101 | 0.06 |

Table 6. P-value table for Table 4: associations between types of data use and needed data type. Significance was determined at the p<.05 level with a Bonferroni correction with m= 70. Significant associations are marked with an asterisk and colored in blue. "Other" options are not shown as there were no significant associations present.

|  | New study | New project | New ideas | Integration | New methods | Inputs | Calibration | Benchmark | Verification | Trends | Comparison | Summ/vis/tools |
|---|---|---|---|---|---|---|---|---|---|---|---|---|
| **New project** | p < .001* | | | | | | | | | | | |
| **New ideas** | p < .001* | p < .001* | | | | | | | | | | |
| **Integration** | p < .001* | p < .001* | p < .001* | | | | | | | | | |
| **New methods** | p < .001 | p < .001* | p < .001* | p < .001* | | | | | | | | |
| **Inputs** | 0.097 | 0.245 | 0.028 | p < .001* | p < .001* | | | | | | | |
| **Calibration** | 0.31 | 0.157 | 1.000 | 0.001 | p < .001* | p < .001* | | | | | | |
| **Benchmark** | 0.148 | 0.133 | 0.764 | 0.003 | p < .001* | p < .001* | p < .001* | | | | | |
| **Verification** | 0.001 | p < .001* | p < .001* | p < .001 | p < .001* | 0.92 | p < .001* | p < .001* | | | | |
| **Trends** | p < .001* | p < .001* | p < .001* | p < .001* | p < .001* | p < .001* | 0.003 | p < .001* | 0.001 | | | |
| **Comparison** | p < .001* | p < .001* | p < .001* | p < .001* | p < .001* | p < .001* | 0.017 | 0.001 | p < .001* | p < .001* | | |
| **Summ/vis/tools** | 0.013 | p < .001* | p < .001* | p < .001* | p < .001* | p < .001* | 0.001 | p < .001* | 0.072 | p < .001* | p < .001* | |
| **Teaching** | p < .001* | p < .001* | p < .001* | 0.005 | p < .001 | 0.063 | 0.699 | 0.045 | p < .001* | 0.016 | p < .001 | p < .001* |

Table 7. P-value table for Table 4: associations between types of data use and other data uses. Significance was determined at the p<.05 level with a Bonferroni correction with m= 196. Significant associations are marked with an asterisk and colored in blue. "Other" options are not shown as there were no significant associations present; duplicate values were removed.

| | New study | New project | New ideas | Integration | New methods | Inputs | Calibration | Benchmark | Verification | Trends | Comparison | Summ/ vis/tools | Teaching |
|---|---|---|---|---|---|---|---|---|---|---|---|---|---|
| **Agricul** | 0.689 | 0.377 | 0.225 | 0.114 | 0.034 | 0.374 | 0.862 | 0.862 | 0.007 | 0.139 | 0.517 | 0.365 | 0.175 |
| **Arts hum** | 0.083 | 0.186 | 0.086 | 0.045 | 0.024 | 0.008 | 0.08 | 0.185 | 1.000 | 0.532 | 0.42 | 0.442 | 0.001 |
| **Astronom** | 0.31 | 0.264 | 1.000 | 0.067 | 0.417 | 0.007 | p < .001* | 0.112 | 0.327 | 0.008 | 0.016 | 0.348 | 0.104 |
| **Biochem** | 0.68 | 0.052 | p < .001 | 0.371 | p < .001 | 0.806 | 0.192 | 0.21 | p < .001* | 0.48 | 0.001 | 0.549 | 0.862 |
| **Biolog** | 1.000 | 0.029 | 0.038 | p < .001* | 0.065 | 0.114 | 0.104 | 0.002 | p < .001 | 0.04 | p < .001* | 0.699 | 0.488 |
| **Busin** | 0.025 | 0.806 | 0.359 | 0.175 | 0.888 | 0.23 | 0.823 | 0.004 | 0.202 | p < .001* | 0.56 | 0.357 | 0.68 |
| **Chem** | 0.699 | 0.699 | 0.435 | 0.124 | 0.071 | 0.081 | 0.06 | 0.777 | 0.001 | 0.374 | 0.345 | 0.086 | 0.269 |
| **Chem Eng** | 0.467 | 0.462 | 0.578 | 0.484 | 0.001 | 0.051 | p < .001* | 0.383 | 0.002 | 0.216 | 0.129 | 0.92 | 0.708 |
| **Comp Sci** | p < .001* | 0.001 | 0.271 | 0.699 | p < .001* | p < .001* | 0.076 | p < .001* | 0.038 | 0.315 | 0.154 | 0.007 | 0.02 |
| **Decis Sci** | 0.343 | 0.038 | 0.286 | 0.01 | 0.028 | p < .001* | 0.004 | 0.038 | 0.043 | 0.004 | 0.129 | 0.036 | 0.386 |
| **Dentist** | 0.752 | 0.041 | 0.92 | 0.124 | 0.708 | 0.022 | 1.000 | 0.41 | 0.271 | 0.119 | 0.92 | 0.343 | 0.578 |
| **Earth Plan** | 0.091 | 0.718 | 1.000 | 0.001 | 0.063 | p < .001* | p < .001* | 0.337 | 0.791 | 0.003 | p < .001* | 0.001 | 0.92 |
| **Econ** | 0.006 | 1.000 | 0.475 | p < .001 | 0.247 | 0.001 | 0.624 | 0.005 | p < .001 | p < .001 | 1.000 | 0.267 | 0.532 |
| **Energy** | 0.841 | 0.522 | 0.168 | 0.791 | 0.017 | p < .001* | p < .001 | p < .001* | 0.027 | 0.502 | 0.502 | 0.708 | 0.162 |
| **Eng Tech** | 0.032 | 0.028 | 0.043 | 0.002 | 0.001 | p < .001* | p < .001* | p < .001* | 0.299 | 0.862 | 0.191 | 0.299 | 0.007 |
| **Environ** | 0.603 | 0.256 | 0.11 | p < .001* | 0.049 | p < .001* | 0.003 | 0.252 | 0.025 | 0.001 | 0.003 | p < .001* | 0.699 |
| **Health Prof** | 0.699 | 0.003 | p < .001 | 0.003 | 1.000 | 0.108 | 0.68 | 0.841 | 1.000 | 0.118 | 0.639 | 0.125 | 0.617 |
| **Immun** | 0.08 | 0.006 | p < .001 | 0.584 | 0.021 | 0.317 | 0.044 | 0.252 | 0.002 | 1.000 | p < .001 | 0.48 | 0.185 |
| **Info Sci** | 0.343 | 0.371 | 0.322 | p < .001 | 0.089 | p < .001 | 0.121 | 0.001 | 0.708 | p < .001 | 0.104 | p < .001* | 1.000 |
| **Math** | 0.093 | 0.343 | 0.325 | 0.198 | 0.005 | p < .001* | 0.292 | 0.301 | 0.047 | 0.275 | 0.74 | 0.357 | 0.584 |
| **Matl Sci** | 0.647 | 0.374 | 0.34 | 0.012 | 0.006 | 0.488 | 0.179 | 0.493 | p < .001* | 0.24 | 0.048 | 0.271 | 0.427 |
| **Med** | 0.313 | 0.045 | 0.002 | 0.597 | 0.014 | 0.005 | 0.004 | 0.012 | 1.000 | 0.791 | 0.427 | 0.256 | 0.036 |
| **Multi** | 1.000 | 0.002 | 0.044 | p < .001* | 0.034 | 0.002 | 0.027 | 0.031 | 0.699 | p < .001* | p < .001* | p < .001* | 0.022 |
| **Neuro** | 0.823 | 0.306 | 0.238 | 1.000 | 0.071 | 0.791 | 0.752 | 0.279 | 0.045 | 0.806 | 0.362 | 0.021 | 0.174 |
| **Nurs** | 0.45 | 0.026 | 0.126 | 0.671 | 0.13 | 0.046 | 1.000 | 0.584 | 0.357 | 0.566 | 0.777 | 0.224 | 0.17 |
| **Pharma** | 0.823 | 0.431 | 0.017 | 0.493 | 0.009 | 0.192 | 0.92 | 0.68 | 0.039 | 0.137 | 0.035 | 0.105 | 0.439 |
| **Physics** | 0.427 | 0.02 | 0.003 | 0.002 | 0.862 | p < .001* | p < .001* | 0.002 | 0.024 | 0.296 | 0.517 | 0.084 | 0.049 |
| **Psych** | 0.064 | 0.383 | 0.088 | 0.064 | 0.399 | 0.006 | 0.92 | 0.332 | 1.000 | 0.841 | 0.003 | 0.507 | 0.014 |
| **Soc Sci** | p < .001 | 0.001 | 0.11 | 0.004 | 0.001 | p < .001* | p < .001 | 0.512 | 0.023 | 0.043 | 0.036 | 0.037 | 0.015 |
| **Vet** | 0.538 | 0.68 | 0.191 | 0.888 | 0.427 | 0.097 | 0.777 | 0.173 | 0.632 | 1.000 | 0.399 | 0.841 | 0.252 |

Table 8. P-value table for Figure 8: associations between disciplinary domain and data use. Significance was determined at the p<.05 level with a Bonferroni correction with m= 434. Significant associations are marked with an asterisk and colored in blue. "Other" options are not shown as there were no significant associations present.

| | New study | New project | New ideas | Integration | New methods | Inputs | Calibration | Benchmark | Verification | Trends | Comparison | Summ/vis/tools | Teaching |
|---|---|---|---|---|---|---|---|---|---|---|---|---|---|
| **Conditions/ methodology** | 0.001* | p < .001* | 0.001* | 0.138 | 0.225 | 0.125 | 0.089 | 0.218 | p < .001* | 0.07 | 0.016 | 0.1 | 0.016 |
| **Processing/handling** | 0.001* | p < .001* | p < .001* | 0.04 | 0.013 | 0.371 | 0.006 | 0.527 | 0.004* | 0.004* | p < .001* | 0.014 | 0.007 |
| **Topic relevance** | 0.197 | p < .001* | 0.001* | 0.074 | 0.009 | 0.481 | 0.358 | 0.458 | 0.613 | 0.038 | 0.623 | 0.031 | 0.017 |
| **Coverage** | 0.022 | 0.167 | 0.331 | p < .001* | 0.093 | 0.012 | 0.517 | 0.139 | 0.666 | p < .001* | p < .001* | p < .001* | 0.041 |
| **Original purpose** | 0.385 | 0.002* | p < .001* | p < .001* | 0.602 | 0.001* | 0.151 | 0.438 | 0.021 | 0.02 | 0.627 | 0.71 | 0.006 |
| **Metadata/ documentation** | 0.313 | 0.338 | 0.046 | p < .001* | p < .001* | 0.012 | 0.318 | 0.561 | 0.506 | 0.044 | p < .001* | 0.07 | 0.018 |
| **Format** | 0.798 | 0.58 | 0.351 | 0.229 | 0.103 | 0.323 | 0.263 | 0.714 | 0.692 | 0.215 | 0.279 | 0.22 | 0.002* |
| **Size** | 0.949 | 0.42 | 0.238 | 0.58 | 0.159 | 0.082 | 0.486 | 0.738 | 0.706 | 0.113 | 0.281 | 0.69 | 0.013 |
| **Ease of access** | 0.622 | 0.037 | 0.002* | 0.278 | 0.765 | 0.917 | 0.136 | 0.661 | 0.12 | 0.074 | 0.334 | 0.571 | 0.005 |
| **Licensing** | 0.606 | 0.188 | 0.202 | 0.100 | 0.078 | 0.146 | 0.088 | 0.087 | 0.916 | 0.297 | 0.369 | 0.004 | 0.001* |
| **Source reputation** | 0.383 | 0.007 | p < .001* | 0.332 | 0.179 | 0.582 | 0.013 | 0.499 | 0.011 | 0.076 | 0.886 | 0.387 | 0.001* |
| **Creator reputation** | 0.049 | 0.017 | 0.035 | 0.368 | 0.654 | 0.565 | 0.383 | 0.475 | 0.029 | 0.01 | 0.523 | 0.358 | 0.042 |
| **Knowing creator** | 0.377 | 0.427 | 0.117 | 0.201 | 0.051 | 0.226 | 0.666 | 0.835 | 0.329 | 0.015 | 0.634 | 0.772 | 0.717 |

Table 9. P-value table for Figure 15: associations between data use and evaluation criteria. Significance was determined at the p<.05 level with a Bonferroni correction with m= 196. Significant associations are marked with an asterisk and colored in blue. "Other" options are not shown as there were no significant associations present.

# Appendix C: Sources used in disciplinary subset

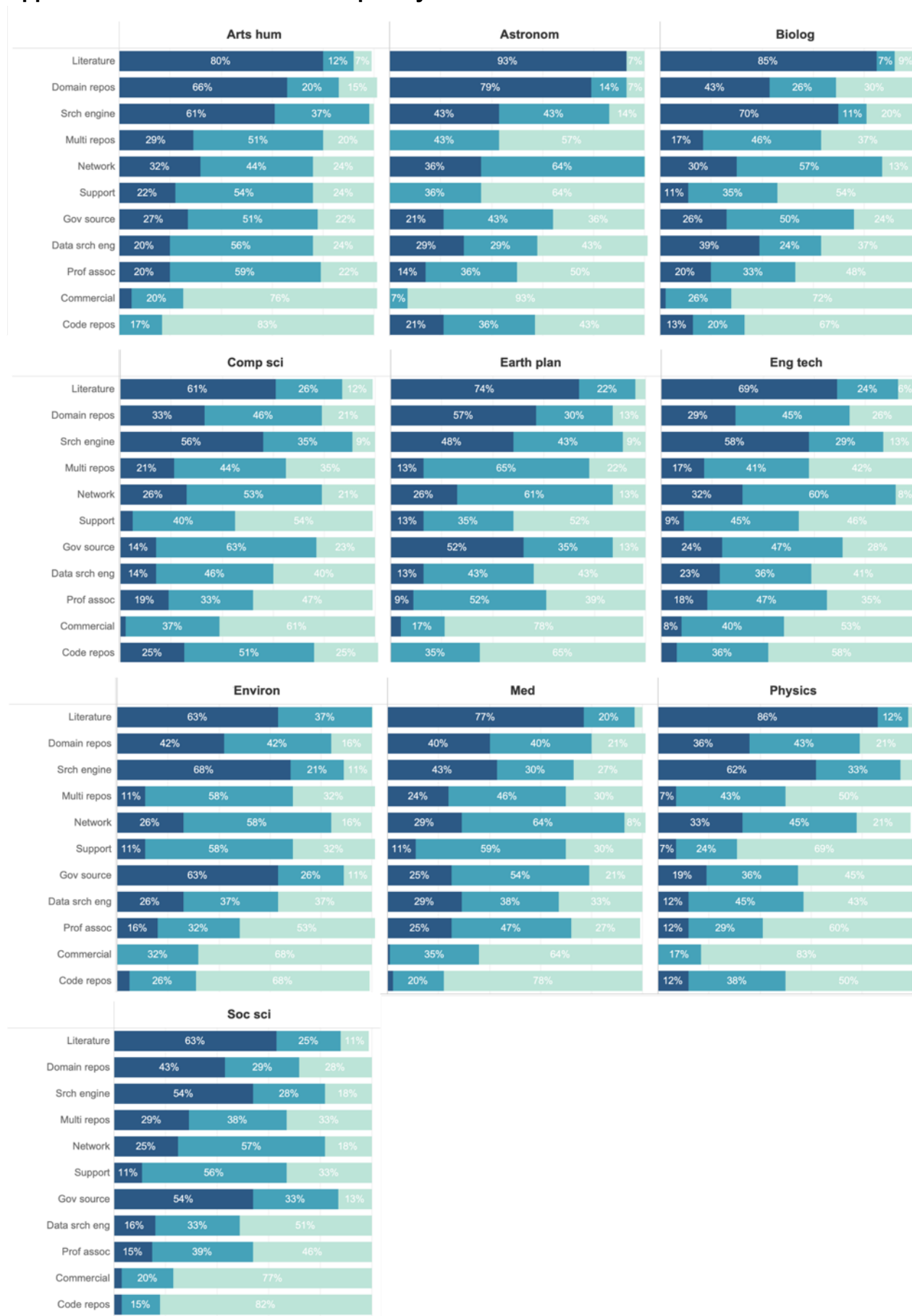

Figure 17. Sources used in the disciplinary subset for respondents selecting only one discipline. Percents are percent respondents. Arts & humanities (n=43); astronomy (n=14); biological science (n=46); computer science (n=57); earth & planetary science (n=24); engineering & technology (n=80); environmental science (n=22); medicine (n=91); physics (n=42); social science (n=81).